\documentclass[twocolumn,pra,superscriptaddress,showpacs]{revtex4}
\usepackage{graphicx}
\usepackage{dcolumn}
\usepackage{amsmath}
\usepackage{color}
\newcommand{\Ninej}[9]{\left\{\begin{matrix}
 {#1} & {\!\!\!#2} & {\!\!\!#3}\cr
 {#4} & {\!\!\!#5} & {\!\!\!#6}\cr
 {#7} & {\!\!\!#8} & {\!\!\!#9}\cr \end{matrix}
\right\}}

\newcommand{\Cleb}[6]{C^{{\,#1}{\,#2}{\,#3}}
  _{{\,#4}{\,#5}{\,#6}} }

\newcommand{\Sixj}[6]{
\left\{\begin{matrix}
 {#1} & {\!\!\!#2} & {\!\!\!#3}\cr
 {#4} & {\!\!\!#5} & {\!\!\!#6}\cr \end{matrix}
\right\}}

\newcommand{\etal}{\textit{et al.}}

\newcolumntype{N}{D..{3.3}}
\newcolumntype{Z}{D..{2.3}}

\begin{document}

\title{Ultracold homonuclear and heteronuclear collisions in metastable helium}
\author{Daniel G. Cocks}
\affiliation{Research School of Science and Engineering, Australian National University, Canberra, Australia 0200}
\affiliation{College of Science and Engineering,
James Cook University, Townsville, Australia 4811}
\author{Ian B. Whittingham}
\affiliation{College of Science and Engineering,
James Cook University, Townsville, Australia 4811}
\author{Gillian Peach}
\affiliation{Department of Physics and Astronomy, University College London,
London WC1E 6BT, UK}

\date{\today}

\begin{abstract}
  Scattering and ionizing cross sections and rates are calculated for ultracold
  collisions between metastable helium atoms using a fully quantum-mechanical
  close-coupled formalism. Homonuclear collisions of the bosonic
  ${}^{4}$He$^{*} +{}^{4}$He$^{*}$ and fermionic
  ${}^{3}$He$^{*} + {}^{3}$He$^{*}$ systems, and heteronuclear collisions of the
  mixed ${}^{3}$He$^{*} +{}^{4}$He$^{*}$ system, are investigated over a
  temperature range 1~$\mu$K to 1~K.  Carefully constructed Born-Oppenheimer
  molecular potentials are used to describe the electrostatic interaction
  between the colliding atoms, and complex optical potentials used to represent
  loss through ionization from the ${}^{1,3}\Sigma $ states. Magnetic spin-dipole
  mediated transitions from the ${}^{5}\Sigma $ state are included and results
  reported for spin-polarized and unpolarized systems. Comparisons are made with
  experimental results, previous semi-classical models, and a perturbed single
	channel model.
\end{abstract}

\maketitle

\section{Introduction}

Knowledge of the dynamics of ultracold collisions in dilute quantum gases is
crucial to the understanding of the cooling and trapping of these gases. The
precision and control of these gases allows for the investigation of many-body
phenomena in quantum degenerate gases \cite{Bloch2008} such as quantized
vortices \cite{Tsubota2013} and topological states in optical lattices
\cite{Cooper2019}. Metastable rare gases are of particular interest as the
release of the large internal energy can be used by experimentalists to easily
detect individual events with high resolution using a microchannel plate and to
potentially count each atom which has been ionized or has escaped from the trap
\cite{Vassen2012}.

Metastable helium is an attractive prospect for experimental and theoretical
studies of fundamental aspects of ultracold collisions because it has only one
active electron, accurate molecular potentials exist to represent the
interaction between the colliding atoms, and large numbers ($\agt 10^{8}$) of
both the bosonic ${}^{4}$He$^{*} \equiv {}^{4}$He(1s2s\;${}^{3}$S$_{1}$) and
fermionic ${}^{3}$He$^{*} \equiv {}^{3}$He(1s2s\;${}^{3}$S$_{1}$) isotopes can
be trapped~\cite{Stas2004}, allowing the investigation of the effects of
different atomic structures and quantum statistical symmetries. Previous studies
have successfully demonstrated the Hanbury-Brown-Twiss effect for both fermionic
and bosonic degenerate gases \cite{Jeltes2007}, ghost imaging with correlated
atom pairs \cite{Khakimov2016} and have tested quantum electrodynamic
calculations through precise tune-out wavelength measurement \cite{Henson2015}.

The ionizing processes
\begin{equation}
\label{R1}
\mathrm{He}^{*} + \mathrm{He}^{*} = \left\{ \begin{array}{ll}
		\mathrm{He} + \mathrm{He}^{+} + e^{-}, & (\mathrm{PI}) \\
		\mathrm{He}^{+}_{2} + e^{-}, & (\mathrm{AI}) 
		\end{array}  \right.
\end{equation}	
where PI stands for Penning ionization and AI for associative ionization, are an
important source of loss of trapped atoms. As the detailed mechanisms involved
are not important in the present study we shall use PI to denote both processes.

As the ${}^{3}$He$^{*}$ and ${}^{4}$He$^{*}$ metastable atoms both have an
electronic spin of $s=1$, these ionization processes are suppressed for an
incoming state with total spin $S=2$ since they would violate spin
conservation. The very weak spin-dipole magnetic interaction can produce spin
flips and mediate PI in collisions with $S=2$ but the corresponding ionization
rate is four orders of magnitude less than that for collisions with $S=0$ or
$S=1$ for which the total electronic spin is
conserved~\cite{Venturi2000,Sirjean2002}.

Homonuclear ionizing collisions of the bosonic ${}^{4}$He$^{*} +{}^{4}$He$^{*}$
system have been investigated experimentally by Mastwijk \textit{et
  al.}~\cite{Mastwijk1998}, Tol \textit{et al.}~\cite{Tol1999}, Kumakura and
Morita~\cite{KM1999}, and Stas \textit{et al.}~\cite{Stas2006}.  The measured
unpolarized ionization rates $K({}^{4}\text{He})$ differed significantly between
the various groups. Kumakura and Morita~\cite{KM1999}, and Stas \textit{et
  al.}~\cite{Stas2006} have also studied collisions in the fermionic
${}^{3}$He$^{*} + {}^{3}$He$^{*}$ system but their measured rates
$K({}^{3}\text{He})$ differ widely.  Both Kumakura and Morita~\cite{KM1999}, and
Stas \textit{et al.}~\cite{Stas2006}, proposed simple semi-classical models in
which the inelastic scattering is viewed as a two-stage process of scattering
from the molecular potential $V(R)$ at large internuclear distance
($R \agt 100 \,a_{0}$) and ionization at small internuclear distance
($R \simeq 5\, a_{0}$). Since the spin-dipole interaction is ignored, the ionization
probability ${}^{2S+1}P_{\text{ion}}$ is assumed to be zero for $S=2$ and is
taken to be unity for $S=0,1$.

The semi-classical models differ in their calculation of the probability that
the colliding atoms reach the distance at where ionization occurs. Kumakura and
Morita ignore tunneling of each partial wave through its centrifugal barrier and
assume the evolution of the scattering states can be approximated by an
adiabatic transition in order to derive the number of accessible ionization
channels. Stas \textit{et al.} calculate the tunneling probabilities and find
considerable quantum reflection for $s$-wave scattering, even though there is no
centrifugal barrier, due to the mismatch between the large wavelength asymptotic
de Broglie wave and the rapidly oscillating wave at small $R$.  They also find
the system is well approximated by a diabatic transition between the long-range
atomic states and short-range molecular states.  The two theoretical models give
quite different results with the Stas \textit{et al.}  model in good agreement
with their experimental results for ionization rates in the bosonic and
fermionic systems. In section \ref{sec:results} we will point out that
considering only a two-stage process neglects an important contribution to the
ionization rate and that the comparison between experiment and theory is
complicated by the mixture of trapped states.

For the bosonic case a detailed theoretical study of elastic, inelastic and
ionization rates using a fully quantum-mechanical close-coupling calculation
already existed~\cite{Venturi2000,Leo2001}. The ionization rates from the Stas
\textit{et al.} model were in moderate agreement with those of this multichannel
calculation.

For the heteronuclear mixed ${}^{3}$He$^{*} +{}^{4}$He$^{*}$ system, McNamara
\textit{et al.}~\cite{McNamara2007} have measured the ionization rate and
extended the Stas \textit{et al.} theoretical model to this system. They
undertook a comparison of the bosonic, fermionic and mixed systems, and found
the experimental results and theoretical model to be in good agreement.

The Stas \textit{et al.} theoretical model has been revisited by
Dickinson~\cite{Dickinson2007} who showed that the stage of quantum reflection
from the molecular potential can be modelled analytically for cold-atom
collisions purely in terms of the long-range van der Waals coefficient and the
particle masses. Ionization rates for unpolarized beams of bosonic, fermionic
and mixed systems of metastable helium atoms obtained from the two models agreed
well over the temperature range from 1~$\mu$K to 2~mK.

Detailed studies of ultracold collisions of metastable helium require fully
quantum-mechanical methods because the onset of quantum threshold behavior
cannot be described semiclassically~\cite{Julienne1989}.  We report here an
extension of our earlier calculations for the bosonic
system~\cite{Venturi2000,Leo2001} to the fermionic and mixed systems.  We
calculate scattering and ionizing cross sections and rates over a temperature range
1~$\mu$K to 1~K using carefully constructed Born-Oppenheimer molecular potentials
and complex optical potentials to represent loss through ionization.

The paper is organized as follows. In Sec. II the theoretical formalism
describing the collisions of the metastable helium atoms is presented. The
close-coupled scattering equations are derived, the molecular basis states
appropriate to the various systems discussed and explicit expressions obtained
for the Hamiltonian matrix elements. The calculation of cross sections and
transition rate coefficients for scattering and ionizing collisions and the
extraction of the required scattering matrix elements from the asymptotic
solutions of the close-coupled equations are then discussed.  In Sec.~\ref{sec:single-channel}
we also discuss a
simple perturbed single-channel model.  The results of our calculations are presented and
discussed in Sec.~\ref{sec:results}, and a summary of the outcomes of this investigation is
given in Sec.~\ref{sec:summary}. Further details of the evaluation of the Hamiltonian matrix
elements and the numerical solution of the multichannel equations are provided
in Appendices A and B respectively.

Atomic units are used, with lengths in Bohr radii $a_{0}=0.0529177209$ nm and energies in
Hartree $E_{\text{h}}= \alpha^{2}m_{e}c^{2}=27.211384$ eV.

\section{Theory}
\subsection{Multichannel equations}

The total Hamiltonian for the system of two interacting metastable helium atoms $i=1,2$ 
with reduced mass $\mu$, interatomic separation $R$ and relative angular momentum 
$\hat{\mathbf{l}}$, is
\begin{equation}
\label{R2}
\hat{H} = \hat{T}_{K}+\hat{H}_{\text{rot}} + \hat{H}_{\text{el}} +\hat{H}_{\text{hfs}} 
+ \hat{H}_{\text{sd}},
\end{equation}
where $\hat{T}_{K}$ is the radial kinetic energy operator
\begin{equation}
\label{R3}
\hat{T}_{K} = -\frac{\hbar^{2}}{2 \mu R^{2}} \frac{\partial}{\partial R} 
\left( R^{2} \frac{\partial }{\partial R} \right),
\end{equation}
and $\hat{H}_{\text{rot}}$ is the rotational operator
\begin{equation}
\label{R4}
\hat{H}_{\text{rot}} = \frac{\hat{l}^{2}}{2 \mu R^{2}}.
\end{equation}
The total electronic Hamiltonian is
\begin{equation}
\label{R5}
\hat{H}_{\text{el}} = \hat{H}_{1} + \hat{H}_{2} + \hat{H}_{12}
\end{equation}
where $\hat{H}_{i}$ is the unperturbed Hamiltonian of atom $i$ and $\hat{H}_{12}$ is the
electrostatic interaction between the atoms. The term $\hat{H}_{\text{hfs}}$ describes the
hyperfine structure of the ${}^{3}$He$^{*}$ atom and must be included for the 
${}^{3}$He$^{*} + {}^{3}$He$^{*}$ and ${}^{3}$He$^{*} +{}^{4}$He$^{*}$ systems. The 
spin-dipole magnetic interaction between the atoms is
\begin{equation}
\label{R6}
\hat{H}_{\text{sd}} = - \frac{\xi }{\hbar^{2}R^{3}} 
\left[ 3 (\hat{\mathbf{S}}_{1}\cdot \hat{\mathbf{R}}) (\hat{\mathbf{S}}_{2}\cdot \hat{\mathbf{R}}) 
- \hat{\mathbf{S}}_{1} \cdot \hat{\mathbf{S}}_{2} \right]
\end{equation}
where $\hat{\mathbf{S}}_{i}$ are the electronic-spin operators, $\hat{\mathbf{R}}=\mathbf{R}/R$
is a unit vector directed along the internuclear axis, and
\begin{equation}
\label{R7}
\xi = \alpha^{2} \left( \frac{\mu_{e}}{\mu_{B}}\right)^{2} \;E_{\text{h}} a_{0}^{3}.
\end{equation}
Here $(\mu_{e}/\mu_{B})=1.00115965 $ is the ratio of the electron magnetic moment to the Bohr 
magneton. 

The multichannel equations describing the interacting atoms are obtained by expanding the system 
eigenstate $|\Psi \rangle $, which satisfies
\begin{equation}
\label{R8}
\hat{H} |\Psi \rangle = E |\Psi \rangle ,
\end{equation}
as
\begin{equation}
\label{R9}
|\Psi \rangle = \sum_{a} \frac{1}{R} G_{a}(R) |a \rangle  ,
\end{equation}
where $G_{a}(R)$ are radial wave functions and the molecular basis is 
$|a \rangle = |\Phi_{a}(R,q)\rangle $, where $q$ denotes the interatomic polar coordinates 
$(\theta, \phi )$ and electronic coordinates $\{\mathbf{r}_{i}\}$. The state label, $a$,
denotes the set of approximate quantum numbers describing the electronic-rotational states
of the molecule. We make the Born-Oppenheimer (BO) approximation that the basis states 
$|a \rangle $ depend only parametrically on $R$ so that 
$\langle a^{\prime}|\hat{T}_{K}|a \rangle = 0$. Forming the scalar product
$\langle a^{\prime}|\hat{H} |a \rangle $ yields the set of multichannel equations
\begin{equation}
\label{R10}
\sum_{a} \left[ - \frac{\hbar^{2}}{2 \mu } \frac{d^{2}}{dR^{2}}\delta_{a^{\prime},a}
+V_{a^{\prime}a}(R)-E\delta_{a^{\prime},a} \right] G_{a}(R) =0,
\end{equation}
where
\begin{equation}
\label{R11}
V_{a^{\prime}a}(R)=\langle a^{\prime}|\left[\hat{H}_{\text{rot}} + \hat{H}_{\text{el}}
+\hat{H}_{\text{hfs}} + \hat{H}_{\text{sd}} \right] |a \rangle .
\end{equation}

\subsection{Basis states and matrix elements}

The molecular basis states $\{|a \rangle \}$ must be chosen such that, in the limit 
$R \rightarrow \infty$, they diagonalize the non-interacting two-atom system. As both
${}^{3}$He$^{*}$ and ${}^{4}$He$^{*}$ have zero orbital angular momentum and only 
${}^{3}$He$^{*}$ has a nuclear angular momentum $\hat{i}$, the appropriate coupling schemes
are
\begin{equation}
  \hat{\mathbf{S}} = \hat{\mathbf{S}}_{1}+\hat{\mathbf{S}}_{2}
\end{equation}
for ${}^{4}$He$^{*} +{}^{4}$He$^{*}$, 
\begin{equation}
\hat{\mathbf{f}}_{1}=\hat{\mathbf{S}}_{1}+\hat{\mathbf{i}}_{1}, \quad 
\hat{\mathbf{f}}_{2}=\hat{\mathbf{S}}_{2}+\hat{\mathbf{i}}_{2}, \quad
\hat{\mathbf{f}}=\hat{\mathbf{f}}_{1}+\hat{\mathbf{f}}_{2}
\end{equation}
for ${}^{3}$He$^{*} +{}^{3}$He$^{*}$, and 
\begin{equation}
\hat{\mathbf{f}}_{1}=\hat{\mathbf{S}}_{1}+\hat{\mathbf{i}}_{1}, \quad
\hat{\mathbf{f}}=\hat{\mathbf{f}}_{1}+\hat{\mathbf{S}}_{2}
\end{equation}
for ${}^{3}$He$^{*} +{}^{4}$He$^{*}$, where, in this last case, we have labelled the 
${}^{3}$He$^{*}$ and ${}^{4}$He$^{*}$ atoms as atom 1 and 2 respectively. Hereafter 
we shall denote the three cases as 4--4, 3--3 and 3--4.

The space-fixed eigenstates for the 3--3 system are
\begin{equation}
\label{R12}
|S_{1},i_{1},f_{1}, S_{2}, i_{2},f_{2}, f, m_{f} \rangle ,
\end{equation}
which simplify to
\begin{equation}
\label{R13}
|S_{1},i_{1}, f_{1}, S_{2}, f, m_{f} \rangle
\end{equation}
for the 3--4 system, and to
\begin{equation}
\label{R14}
|S_{1}, S_{2}, S,M_{S} \rangle 
\end{equation}
for the 4--4 system. Here $m_{j}$ denotes the projection of an angular momentum 
$\hat{\mathbf{j}}$ onto the space-fixed $Oz$ quantization axis.

We denote these states generically by $|\Gamma, f, m_{f} \rangle $, where
$\Gamma = \{\alpha_1, \alpha_2\}$ and $\alpha_i = \{S_{i},i_{i},f_{i}\}$, with
the simplifications $\alpha_{2}=S_{2}$ for the 3--4 system and, for the 4--4
system, $\alpha_{i}=S_{i}$ and $(f,m_{f})=(S,M_{S})$.  Although the desired
cross sections will be expressed in terms of the states
\begin{equation}
\label{R15}
|\beta \rangle \equiv |\Gamma, f, m_{f}, l, m_{l} \rangle = |\Gamma, f, m_{f} \rangle |l, m_{l} \rangle ,
\end{equation}
where $|l, m_{l} \rangle $ are the relative motion eigenstates, it is more
convenient to perform the calculations with the coupled states
\begin{equation}
\label{R16}
|\Gamma, f, l, J, M_{J} \rangle = \sum_{m_{f},m_{l}} C^{flJ}_{m_{f}m_{l}M_{J}}
|\Gamma, f, m_{f} \rangle |l, m_{l} \rangle ,
\end{equation}
where $\hat{\mathbf{J}}=\hat{\mathbf{f}}+\hat{\mathbf{l}}$ is the total angular
momentum.  This simplifies the calculations as $J$ and $M_{J}$ are conserved and
fewer coupled equations are required since they are independent of $M_{J}$.
In (\ref{R16}) \mbox{\footnotesize{$\Cleb{j_{1}}{j_{2}}{j}{m_{1}}{m_{2}}{m}$}}
is a Clebsch-Gordan coefficient.

The states $|\Gamma,f,l,J,M_J \rangle$ with $i_1$,$i_2$ arbitrary are not symmetrized under 
$\hat{X}_N$, the operator that permutes the nuclear labels.
We form the symmetrized states 
\begin{eqnarray}
\label{R16-x}
|a\rangle & \equiv & |\Gamma ; f,l,J,M_J,X_N\rangle    \nonumber\\
& = & N_{X_N} \left[|a_{12} \rangle + (-1)^{X_N+l+f_1+f_2-f}|a_{21} \rangle \right],
\end{eqnarray}
where
\begin{equation}
\label{R16a}
|a_{12}\rangle = |\left(\alpha_1\right)_A,\left(\alpha_2\right)_B,f,l,J,M_J\rangle ,
\end{equation}
and
\begin{equation}
\label{R16b}
|a_{21}\rangle = |\left(\alpha_2\right)_A,\left(\alpha_1\right)_B,f,l,J,M_J\rangle .
\end{equation}
Here, the subscripts $A$ and $B$ indicate the labelling of the nuclei, the
normalization constant is $N_{X_N}=1/\sqrt{2(1+\delta_{\alpha_{1},\alpha_{2}})}$,
and $X_{N}=0(1)$ for bosonic (fermionic) systems.  This gives the selection rule
$(-1)^{l-S} = 1$ for the 4--4 system and $(-1)^{l+f_1+f_2-f} = 1$ for the 3--3
system.  For the 3--4 system, the symmetrized states present no advantage and so
we work in the unsymmetrized basis, with $i_1 = \frac{1}{2}$ and $i_2 = 0$.  We
note that, in the case of the 4--4 system, the selection rule also enforces a
symmetry of $l$ even (odd) for gerade (ungerade) states when considering the
symmetry under electronic inversion.

The multichannel equations (\ref{R10}) require the matrix elements of 
$\hat{H}_{\text{rot}}$, $\hat{H}_{\text{el}}$, $\hat{H}_{\text{hfs}}$ and
$\hat{H}_{\text{sd}}$ in the basis $|a\rangle$. The rotation matrix elements 
are simply 
\begin{equation}
\label{R17}
\langle a^{\prime} |\hat{H}_{\text{rot}}  | a \rangle  =
\frac{\langle a^{\prime} |\hat{l}^{2}| a \rangle }{2 \mu R^{2}} = 
\delta_{a, a^{\prime}} \frac{l (l+1) \hbar^{2}}{2 \mu R^{2}}.
\end{equation}

The eigenstates of $\hat{H}_{\mathrm{el}}$ are the body-fixed states arising from 
the coupling $\hat{\mathbf{S}}=\hat{\mathbf{S}}_{1}+\hat{\mathbf{S}}_{2}$ and must
also be eigenstates of the electron inversion operator $\hat{P}_{S}$. They satisfy
\begin{eqnarray}
\label{R18}
\hat{H}_{\text{el}}
|\left(S_{1}\right)_{A}, \left(S_{2}\right)_{B},S, \Omega_{S},w \rangle  = 
{}^{2S+1}V_{\Sigma_{w}}(R)  \nonumber  \\
\times |\left(S_{1}\right)_{A}, \left(S_{2}\right)_{B},S,\Omega_{S} ,w\rangle 
\end{eqnarray}
where $\Omega_{S}$ denotes the projection of $\hat{\mathbf{S}}$ onto the internuclear axis 
$\mathbf{R}$, ${}^{2S+1}V_{\Sigma_{w}}(R)$ are the Born-Oppenheimer molecular
potentials, and $w=0(1)$ for \textit{gerade} (\textit{ungerade}) symmetry. 
The related space-fixed states
\begin{eqnarray}
\label{R18a}
|\left(S_{1}\right)_{A}, \left(S_{2}\right)_{B},S, M_{S}, w \rangle  = 
\sum_{\Omega_{S}} D^{S}_{\Omega_{S},M_{S}}(\phi,\theta,0)  \nonumber  \\
\times |\left(S_{1}\right)_{A}, \left(S_{2}\right)_{B},S, \Omega_{S}, w \rangle ,
\end{eqnarray}
where $D^{S}_{\Omega_{S},M_{S}}(\phi,\theta,0)$ is the Wigner rotation matrix, are also eigenstates
of $\hat{H}_{\text{el}}$, satisfying (\ref{R18}), as the Born-Oppenheimer potentials are 
independent of $\Omega_{S}$.
We note that these states $|\left(S_1\right)_{A},\left(S_2\right)_{B},S,M_S,w\rangle$ are also 
eigenstates of $\hat{P}_{S}$ and are \textit{gerade} (\textit{ungerade})  for even (odd) $S$. 
Hereafter we shall omit the label $w$ on the potentials.
The matrix elements $\langle a^{\prime}|\hat{H}_{\text{el}}|a \rangle $ are constructed
from the matrix elements~\cite{Cocks2015} (see Appendix A)

\begin{eqnarray}
\label{R18b}
\langle a_{12}^\prime | \hat{H}_\text{el} | a_{12} \rangle & = & \delta_{\eta^{\prime},\eta}
			[f_1^\prime f_2^\prime f_1 f_2]^{1/2} \; \sum_{S,i} [Si] \\
					& & \times  \Ninej{S_1}{S_2}{S}{i_1}{i_2}{i}{f_1^\prime}{f_2^\prime}{f}
									\Ninej{S_1}{S_2}{S}{i_1}{i_2}{i}{f_1}{f_2}{f} 			
												{}^{2S+1}V_\Sigma (R), \nonumber
\end{eqnarray}
and the elements $\langle a_{21}^\prime | \hat{H}_\text{el} | a_{21} \rangle $ and 
$\langle a_{21}^\prime | \hat{H}_\text{el} | a_{12} \rangle $ which are obtained 
from (\ref{R18b}) by the obvious replacements. Here $[ab\ldots ]=(2a+1)(2b+1)\ldots $, the notation
\mbox{\footnotesize{$\Ninej{a}{b}{c}{d}{e}{f}{g}{h}{i}$}} is a Wigner 9-j symbol, and
$\eta$ denotes the set of quantum numbers $\{S_{1}, i_{1}, S_{2}, i_{2},f,l,J, M_{J}\}$.
For the 3--4 system, $i_{2}=0$, $f_{2}=f_{2}^{\prime}=S_{2}$ and (\ref{R18b}) 
simplifies to
\begin{eqnarray}
\label{R18d}
\langle a_{12}^\prime | \hat{H}_\text{el} | a_{12} \rangle & = & \delta_{\xi^{\prime},\xi}
(-1)^{f_{1}-f_{1}^{\prime}} [f_{1}^{\prime}f_{1}]^{1/2}  \sum_{S} [S] \\
&& \times \Sixj{S_{1}}{S_{2}}{S}{f}{i_{1}}{f_{1}^{\prime}}
\Sixj{S_{1}}{S_{2}}{S}{f}{i_{1}}{f_{1}} {}^{2S+1}V_\Sigma (R), \nonumber
\end{eqnarray}
where \mbox{\footnotesize{$\Sixj{a}{b}{c}{d}{e}{f}$}} is a 6-j symbol and
$\xi $ denotes the set $\{S_{1},i_{1},S_{2},f,l,J,M_{J}\}$. 
Further simplification occurs for the 4--4 system. 
With $i_{1}=0, f_{1}=f_{1}^{\prime} =S_{1}, f=S$, (\ref{R18b}) reduces to
\begin{equation}
\label{R18c}
\langle a_{12}^\prime | \hat{H}_\text{el} | a_{12} \rangle = \delta_{\rho^{\prime},\rho} 
{}^{2S+1}V_\Sigma(R),
\end{equation}
where $\rho=\{\alpha_{1},\alpha_{2},f,l,J,M_{J}\}$. Note that, for the 3--3 and 3--4 systems, 
$\hat{H}_\text{el} $ couples the different hyperfine levels $f_{i}$.

The matrix elements of $\hat{H}_\text{hfs}$ are assumed to be independent of the
interatomic spacing, such that 
$\hat{H}_\text{hfs} = \hat{H}_\text{hfs,1} + \hat{H}_\text{hfs,2}$, 
where the individual atomic hyperfine splitting matrix elements are independent
of $m_{f_{i}}$ and are given by
\begin{align}
\langle \alpha_i^\prime| \hat{H}_\text{hfs,i} |\alpha_i\rangle =  
\delta_{\alpha_i^\prime,\alpha_i} E^{\text{hfs}}_{i_i,f_i}.
\end{align}
For helium-4, $i_i=0$ and $E^{\text{hfs}}_{0,f_{i}} = 0$. For helium-3 
with hyperfine splitting 
$\epsilon_{\text{hfs}}=6739.701177$~MHz $=1.519830 \times 10^{-7}\,
E_{\text{h}}$~\cite{Rosner1970}, we choose our energy origin on the 
lower hyperfine level such that 
$E^{\text{hfs}}_{0.5,0.5}=\epsilon_{\text{hfs}}$ and $E^{\text{hfs}}_{0.5,1.5}=0$.

The spin-dipole interaction may be written as the scalar product of two
second-rank irreducible tensors 
\begin{equation}
\label{R19}
\hat{H}_{\text{sd}} = V_p(R) \mathbf{T}^2 \mathbf{\cdot} \mathbf{C}^2
\end{equation}
where $\mathbf{T}^2$ is
\begin{equation}
\label{R20}
T^2_q \equiv \left[\mathbf{S}^1_1 \mathbf{\times}\mathbf{S}^1_2 \right]^2_q = 
\sum_\mu \Cleb{1}{1}{2}{\mu,}{q-\mu,}{q}\;S^1_{1,\mu} S^1_{2,q-\mu},
\end{equation}
and $\mathbf{C}^2$ is the second-rank tensor formed from the
modified spherical harmonics
\begin{equation}
\label{R21}
  C^l_{m_{l}} (\theta,\phi) \equiv \sqrt{\frac{4\pi}{2l+1}} \;
  Y_{lm_{l}} (\theta, \phi)
\end{equation} 
where  $Y_{lm_{l}} (\theta, \phi ) = \langle \theta \phi |lm_{l} \rangle $.
The radial factor is $V_p(R) = b/R^3$ where
$b \equiv -\sqrt{6}\xi/\hbar^2$. The matrix elements of 
$\hat{H}_{\text{sd}}$ in the basis
\begin{equation}
\label{R22} 
|\alpha \rangle = |\left(S_{1}\right)_{A},\left(S_{2}\right)_{B}, S, M_{S} \rangle |l,m_{l}\rangle,
\end{equation}
are \cite{Beams2006}
\begin{equation}
\label{R23}
\langle \alpha^{\prime} |\hat{H}_{\text{sd}}| \alpha \rangle =
V_{p}(R)D_{\alpha^{\prime}\alpha }
\end{equation}
where
\begin{eqnarray}
\label{R24}
D_{\alpha^{\prime}\alpha} & =  &
\delta_{M_{S^{\prime}} + m^{\prime}_{l}, M_S + m_{l}}\;
 (-1)^{M_{S^{\prime}} - M_S} \;
\Cleb{S}{2}{S^{\prime}}{M_S,}{M_{S^{\prime}}- M_S,}
 {M_{S^{\prime}}} \nonumber  \\
&& \times  \Cleb{l}{2}{l^{\prime}}{m_{l}}{m^{\prime}_{l}-m_{l}}
{m^{\prime}_{l}} \; 
\langle \gamma^{\prime},S^{\prime}||\mathbf{T}^2||\gamma, S \rangle
 \langle l^{\prime}||\mathbf{C}^2|| l\rangle
\end{eqnarray}
and $\gamma=\{S_{1},S_{2}\}$.
The reduced matrix elements are given by
\begin{eqnarray}
\label{R25}
\langle \gamma^{\prime}, S^{\prime}||\mathbf{T}^2||\gamma, S \rangle & = &
   \delta_{\gamma^{\prime},\gamma} \;\hbar^{2} 
\sqrt{5 S_1(S_1+1)S_2(S_2+1)}  \nonumber \\
&& \times [S_{1}S_{2}S]^{1/2}
\Ninej{S_{1}}{S_{2}}{S}{1}{1}{2}{S_{1}}{S_{2}}{S^{\prime}} 
\end{eqnarray}
and
\begin{equation}
\label{R26}
\langle l^{\prime}||\mathbf{C}^2||l \rangle =
\left[ \frac{2l+1}{2l^{\prime}+1} \right]^
 {\frac{1}{2}}\; \Cleb{l}{2}{l^{\prime}}{0}{0}{0}.
\end{equation}
The nonzero reduced matrix elements are
$\langle \gamma^{\prime},2||\mathbf{T}^{2}||\gamma ,2\rangle =
\delta_{\gamma^{\prime},\gamma}\hbar^{2} \sqrt{7/3} $, 
$\langle \gamma^{\prime},0||\mathbf{T}^{2}||\gamma ,2\rangle =
-\delta_{\gamma^{\prime},\gamma}\hbar^{2} \sqrt{5/3} $, and
$\langle \gamma^{\prime},1||\mathbf{T}^{2}||\gamma ,1\rangle =
\delta_{\gamma^{\prime},\gamma}\hbar^{2} \sqrt{5/3} $.

The conversion from the $|\alpha\rangle$ basis to the 
$|a_{12}\rangle$ basis gives (see Appendix A)
\begin{equation}
\label{R26a}
\langle a_{12}^{\prime}|\hat{H}_{\mathrm{sd}}|a_{12} \rangle =V_{p}(R) D_{a_{12}^{\prime}a_{12}}
\end{equation}
where
\begin{eqnarray}
\label{R26b}
 D_{a_{12}^{\prime}a_{12}} & = & \delta_{\lambda^{\prime},\lambda}(-1)^{l^{\prime}+J} 
[f_{1}^{\prime}f_{2}^{\prime}f^{\prime}f_{1}f_{2}f l^{\prime}]^{1/2}
\langle l^{\prime}||\mathbf{C}^2||l \rangle  \nonumber  \\
&& \times \Sixj{f}{2}{f^{\prime}}{l^{\prime}}{J}{l} 
\sum_{S^{\prime},S,i}(-1)^{-S^{\prime}-i} [S^{\prime} i][S]^{1/2} \nonumber  \\
&& \times \Sixj{f}{2}{f^{\prime}}{S^{\prime}}{i}{S}
\langle \gamma^{\prime}, S^{\prime}||\mathbf{T}^2||\gamma, S \rangle 
\nonumber  \\
&& \times \Ninej{S_1}{S_2}{S^{\prime}}{i_1}{i_2}{i}
{f_1^\prime}{f_2^\prime}{f^{\prime}}
\Ninej{S_1}{S_2}{S}{i_1}{i_2}{i}{f_1}{f_2}{f},
\end{eqnarray}
and $\lambda =\{S_{1},i_{1},S_{2},i_{2},J,M_{J}\}$.
The elements $\langle a_{21}^\prime | \hat{H}_\text{sd} | a_{21} \rangle $ and 
$\langle a_{21}^\prime | \hat{H}_\text{sd} | a_{12} \rangle $ can be obtained 
from (\ref{R26b}) by the obvious replacements. 
This expression simplifies for the 3--4 system to
\begin{eqnarray}
\label{R26c}
 D_{a_{12}^{\prime}a_{12}} & = & \delta_{\tau^{\prime},\tau}
(-1)^{f_{1}-f_{1}^{\prime}+l^{\prime}+J-i_{1}} 
[f_{1}^{\prime}f^{\prime}f_{1}f l^{\prime}]^{1/2}
\langle l^{\prime}||\mathbf{C}^2||l \rangle  \nonumber  \\
&& \times \Sixj{f}{2}{f^{\prime}}{l^{\prime}}{J}{l} 
\sum_{S^{\prime},S}(-1)^{-S^{\prime}} [S^{\prime}][S]^{1/2} \nonumber  \\
&& \times \Sixj{f}{2}{f^{\prime}}{S^{\prime}}{i_{1}}{S}
\langle \gamma^{\prime}, S^{\prime}||\mathbf{T}^2||\gamma, S \rangle 
\nonumber  \\
&& \times \Sixj{S_{1}}{S_{2}}{S^{\prime}}{f^{\prime}}{i_{1}}{f_{1}^{\prime}}
\Sixj{S_{1}}{S_{2}}{S}{f}{i_{1}}{f_{1}},
\end{eqnarray} 
where $\tau =\{S_{1},i_{1},S_{2},J,M_{J}\}$. Finally, for the 4--4 system, we get
\begin{eqnarray}
\label{R26d}
 D_{a_{12}^{\prime}a_{12}} & = & \delta_{\omega^{\prime},\omega}
(-1)^{l^{\prime}+J+S^{\prime}} 
[S^{\prime}l^{\prime}]^{1/2} \langle l^{\prime}||\mathbf{C}^2||l \rangle  \nonumber  \\
&& \times \Sixj{S}{2}{S^{\prime}}{l^{\prime}}{J}{l} 
\langle \gamma^{\prime}, S^{\prime}||\mathbf{T}^2||\gamma, S \rangle ,
\end{eqnarray} 
where $\omega =\{S_{1},S_{2},J,M_{J}\}$.

\subsection{Cross sections}

For collisions in an atom trap, the scattering cross section for transitions
from an initial state $|\Gamma, f, m_{f} \rangle $ with wave number $k_{\Gamma}$ to final states
$|\Gamma^{\prime},f^{\prime},m_{f}^{\prime} \rangle $ should be averaged over initial directions and
integrated over all final directions~\cite{Julienne1989,Julienne2002}, to give
\begin{equation}
\label{R27}
\sigma (k_\Gamma;\Gamma, f, m_{f} \rightarrow \Gamma^{\prime}, f^{\prime}, m_{f}^{\prime}) 
= \frac{\pi N_{\Gamma}}{k^{2}_{\Gamma}} 
\sum_{l^{\prime},m_{l}^{\prime}} \sum_{l, m_{l}}|T_{\beta^{\prime}\beta}(E)|^{2}
\end{equation}
where the factor $N_\Gamma = 2/N_{X_N}^{2} = 1 + \delta_{\alpha_1,\alpha_2}$ accounts for
normalization of the incoming state, and $T_{\beta^{\prime}\beta}(E)$ are the matrix elements
of the transition operator $\mathbf{T}$ between the states (\ref{R15}).  The
channel $\beta $ must be open, that is the energy $E$ of the state must be greater
than the asymptotic energy
$V_{\Gamma}^{\infty} \equiv E^{\mathrm{hfs}}_{i_{1}f_{1}}+E^{\mathrm{hfs}}_{i_{2}f_{2}}$ so
that $k_{\Gamma}^{2} \equiv 2 \mu [E-V^{\infty}_{\Gamma}]/\hbar^{2} > 0$.  These transition matrix
elements are related to those of the total angular momentum representation
(\ref{R16-x}) by
\begin{equation}
\label{R28}
T_{\beta^{\prime}\beta}(E)= \sum_{J^{\prime},M_{J}^{\prime}}
\sum_{J, M_{J}}\Cleb{f^{\prime}}{l^{\prime}}{J^{\prime}}{m_{f}^{\prime}}
{m_{l}^{\prime}}{M_{J}^{\prime}}
\Cleb{f}{l}{J}{m_{f}}{m_{l}}{M_{J}}
T_{a^{\prime}a}(E).
\end{equation} 
This relationship simplifies as the total angular momentum transition matrix 
elements $T_{a^{\prime}a}(E)$ and scattering matrix elements 
$S_{a^{\prime}a}(E)=\delta_{a^{\prime},a}-T_{a^{\prime}a}(E)$
extracted from the asymptotic solutions of (\ref{R10}) are diagonal 
in $J$ and independent of $M_{J}$, allowing the notation 
$T^{J}_{\Gamma^{\prime}f^{\prime}l^{\prime},\Gamma f l}$ and
$S^{J}_{\Gamma^{\prime}f^{\prime}l^{\prime},\Gamma f l}$.
For the homonuclear 3--3 and 4--4 systems, the matrix elements $T_{a^{\prime}a}$ and 
$S_{a^{\prime}a}$ are understood to be in the symmetrized basis (\ref{R16})
whereas, for the heteronuclear 3--4 system, unsymmetrized states are used. 


Spin-polarized systems with spin-stretched states $S=2$ for the 4--4 system, $f=3$ for 
the 3--3 system, and $f=\frac{5}{2}$ for the 3--4 system, are all in the 
$S=2$ ${}^{5}\Sigma^{+}_{g} $
state from which Penning ionization is not possible. However, the 
spin dipole interaction mediates transitions to the ${}^{1}\Sigma^{+}_{g} $ state from 
which Penning ionization is highly probable at small internuclear separations
$R \alt 7a_{0}$. For non-spin-polarized systems with $S=0,1,2$, Penning ionization from the 
${}^{3}\Sigma^{+}_{u} $ state is also highly probable at small $R$. The loss of flux due to 
Penning ionization can be modelled using complex optical potentials 
${}^{1,3}V_{\Sigma}(R)-i {}^{1,3}\Gamma(R)/2$ where ${}^{2S+1}\Gamma (R)$ is the total 
autoionization width. We consider two forms of the autoionization width; a least-squares
fit $\Gamma_{\text{M}}(R)$ to the tabulated values of M\"{u}ller \textit{et al.}~\cite{Muller1991}, 
and the simpler form $\Gamma_{\text{GMS}}(R)=0.3 \exp (-R/1.086)$~\cite{Garrison73}.

The cross section for Penning ionization requires the transition probability from
the initial state $|\Gamma, f, m_{f} \rangle $ to all possible 
ionization channels
\begin{equation}
\label{R29}
\sigma (k_\Gamma;\Gamma, f, m_{f} \rightarrow \text{PI}) = \frac{\pi N_{\Gamma} }{k_{\Gamma}^{2}} 
\sum_{l,m_{l}}|T(E;\beta \rightarrow \text{PI})|^{2}.
\end{equation}
As the loss of flux due to coupling to these ionization 
channels is represented here by a complex potential, the transition 
matrix element for Penning ionization can be obtained from the
non-unitarity of the calculated scattering matrix:
\begin{equation}
\label{R30}
|S(E;\beta \rightarrow \text{PI})|^{2} = 1- \sum_{\beta^{\prime}}
|S_{\beta^{\prime}\beta}(E)|^{2}.
\end{equation} 

Experimental studies usually involve~\cite{Stas2006,McNamara2007} unpolarized systems consisting 
of atoms colliding in all the possible $|\Gamma, f, m_{f} \rangle $ states. 
The contribution of each collision channel depends on the distribution of the magnetic 
substates and makes comparison of theoretical and experimental results very
difficult (see sections~\ref{sec:experimentally-relevant-rates} and \ref{sec:results}).
In order to obtain some specific results, we consider an unpolarized system in which the 
degenerate magnetic substates $|\Gamma, f, m_{f} \rangle $ are populated 
according to their Boltzmann weighting factor $\exp(-V^\infty_\Gamma/k_B T)$, where $k_{B}$ is 
Boltzmann's constant. 
In this distribution, states that differ only in $m_f$ are equally populated, hence 
the cross section for transitions $(\Gamma,f \rightarrow \Gamma^{\prime},f^{\prime})$, obtained 
by averaging (\ref{R27}) over $m_{f}$ and summing over $m^{\prime}_{f}$, is
\begin{equation}
\label{R31}
\sigma (k_\Gamma;\Gamma, f \rightarrow \Gamma^{\prime}, f^{\prime}) 
= \frac{\pi N_{\Gamma}}{k_{\Gamma}^{2}} \sum_{J,l^{\prime},l}\frac{2J+1}{2f+1}\;
|T^{J}_{\Gamma^{\prime}f^{\prime}l^{\prime},\Gamma f l}(E)|^{2}.  
\end{equation}
In our calculations, we have truncated the $J$ summation at $J=20$ to achieve convergence 
for temperatures $T\leq 1$~K. 
The corresponding ionization cross section is
\begin{eqnarray}
\label{R32}
\sigma (k_\Gamma;\Gamma,f \rightarrow \text{PI}) & = &\frac{1}{2f+1}
\sum_{m_{f}} \sigma (k_\Gamma;\Gamma, f, m_{f} \rightarrow \text{PI})
\nonumber  \\ 
&  =  & \frac{\pi N_{\Gamma}}{k^{2}_{\Gamma}} \sum_{J,l} \frac{2J+1}{2f+1} \nonumber  \\
& & \times \left[ 1-\sum_{\Gamma^{\prime},f^{\prime},l^{\prime}}
|S^{J}_{\Gamma^{\prime}f^{\prime}l^{\prime},\Gamma f l}(E)|^{2}\right].
\end{eqnarray}

The transition rate coefficients at temperature $T$ for each of the cross sections are
\begin{equation}
\label{R33}
K(T;x\rightarrow y) = \langle \sigma(k_\Gamma;x \rightarrow y) v \rangle
\end{equation}
where $v=\hbar k_\Gamma/\mu $ is the relative velocity of the colliding atoms.
The angle brackets denote an average over a normalized 
Maxwellian distribution of velocities 
\begin{equation}
\label{R33a}
f(v) = 4 \pi v^{2} \left( \frac{\mu }{2 \pi k_{B}T} \right)^{3/2} 
\exp \left(-\frac{\mu v^{2}}{2 k_{B} T} \right) .
\end{equation}

\subsection{Experimentally relevant rates}
\label{sec:experimentally-relevant-rates}

Many experiments exploit the spin-suppression of ionization in a spin-polarized
cloud. In these experiments there are three major rates of interest. Foremost is
the ionization rate, which is mediated by the spin-dipole interaction, and
causes trap loss. This is given simply by
\begin{equation}
  \label{stretched_loss}
  K_\mathrm{stretched}^\mathrm{loss}(T) = K(T;\Gamma,f^{\mathrm{max}},f^{\mathrm{max}}\rightarrow\mathrm{PI})
\end{equation}
where $f^{\mathrm{max}}=2,\,\frac{5}{2},\,3$ for the 4--4, 3--4 and 3--3 systems,
respectively. Second is the inelastic rate, also mediated by the spin-dipole
interaction, which reduces the overall spin-polarization of the cloud and leads
to subsequent loss. This is given by
\begin{equation}
  \label{stretched_inelastic}
  K_\mathrm{stretched}^\mathrm{inelastic}(T) = \sum_{f^\prime,m_f^\prime}
  K(T;\Gamma,f^{\mathrm{max}},f^{\mathrm{max}}\rightarrow\Gamma^\prime,f^\prime,m_f^\prime).
\end{equation}
where the sum is over all combinations of $f^\prime,m_f^\prime$ except
$f^\prime=m_f^\prime=f^\mathrm{max}$.  The final rate of interest is the elastic scattering rate
\begin{equation}
  \label{stretched_elastic}
  K_\mathrm{stretched}^\mathrm{elastic}(T) = K(T;\Gamma,f^{\mathrm{max}},f^{\mathrm{max}}
	\rightarrow\Gamma,f^{\mathrm{max}},f^{\mathrm{max}}),
\end{equation}
which dominates the other rates.

In experiments that take place in a magneto-optical-trap or are otherwise in a
mixture of the different atomic states, there is a wide range of different
populations of the collisional states $|\Gamma,f,m_f\rangle$. At cold enough
temperatures, we can assume that only the lowest energy hyperfine state will be
occupied, although we will not make that assumption in our results. In general,
there will be a spatially varying population of the magnetic sublevels, due to
the laser coupling and magnetic fields applied in the trap.

As we would like to present results that are generally applicable, we choose to
describe the rates in an ``unpolarized'' system, in which the occupancy of the
atomic states is in thermal equilibrium, i.e. given by the Boltzmann factor. As
this minimizes the proportion of spin-stretched collisions, we expect that an
unpolarized distribution provides an upper bound to the loss rate. We want to
describe the loss and thermalization rates of such an unpolarized system.

In an atomically separable basis $|\gamma_1\rangle|\gamma_2\rangle$, where 
$\gamma_{i}=\{S_{i},i_{i},f_{i},m_{f_{i}}\}$, the collision rate (i.e. number of
collisions per unit time per volume) for loss processes is given by
\begin{equation}
  r_{\gamma_1,\gamma_2}^{\mathrm{PI}} = K(T,\gamma_1,\gamma_2\rightarrow\mathrm{PI}) 
	\frac{n_{\gamma_1}n_{\gamma_2}}{1 + \delta_{\gamma_1,\gamma_2}}
\end{equation}
where $n_{\gamma_i}$ is the atomic density and the denominator prevents over-counting of 
collision pairs for identical atoms. Analogously we have
\begin{equation}
  r_\zeta^{\mathrm{PI}} = K(T;\Gamma,f,m_{f}\rightarrow\mathrm{PI}) \frac{n_\zeta^{(2)}}{N_\Gamma},
\end{equation}
where $\zeta \equiv \{\Gamma, f, m_{f}\}$, the 2-particle density is
$n_\zeta^{(2)}=\sum_{\gamma_1,\gamma_2}
|\langle\gamma_1,\gamma_2|\zeta\rangle|^2 n_{\gamma_1}n_{\gamma_2}$,
 and the factor $N_\Gamma$ is as defined below equation \eqref{R27}.

With these definitions, we can show that, for an unpolarized sample, we have
\begin{equation}
  \sum_{\zeta}r_\zeta^{\mathrm{PI}} = \sum_{\gamma_1,\gamma_2} r_{\gamma_1,\gamma_2}^{\mathrm{PI}}.
\end{equation}
This allows us to introduce a total loss rate for the unpolarized system
\begin{equation}
\label{unpolar_loss}
\tilde{K}^{\mathrm{PI}}_{\mathrm{unpol}} (T) = \frac{1}{Z} \,\sum_{\Gamma, f}e^{-V^\infty_\Gamma/k_{B}T} (2f+1)
\frac{K(T;\Gamma,f \rightarrow \text{PI})}{N_\Gamma}, 
\end{equation}
where the partition function is $Z = \sum_{\Gamma, f}e^{-V^\infty_\Gamma/k_{B}T}(2f+1)$. We note that
$\tilde{K}$ differs to $K$ in that the collision rate is defined purely
quadratically, i.e.
$r^\mathrm{PI}_\mathrm{unpol} = \tilde{K}^\mathrm{PI}_\mathrm{unpol} n^2$ where
$n$ is the total density without any additional factor of 1/2.

In a similar fashion, we can define the thermalization collision rate, i.e. the rate of
collisions which do not change the hyperfine energy of the atomic pair:
\begin{equation}
\label{unpolar_thermal}
\tilde{K}^{\text{thermal}}_{\text{unpol}} (T) = \frac{1}{Z}
\sum_{\Gamma,f,f^\prime} e^{-V^\infty_\Gamma / k_{B}T} (2f+1)
\frac{K(T;\Gamma, f \rightarrow \Gamma, f^\prime)}{N_\Gamma}.
\end{equation}

\subsection{Extraction of $S$-matrix elements}

The $S$-matrix elements required for evaluating the cross sections (\ref{R31}) and (\ref{R32})
are determined by matching the asymptotic solutions of (\ref{R10})
to the combination~\cite{Mies1980} 
\begin{equation}
\label{R34}
\mathbf{G}(R) \underset{R\rightarrow\infty}{\sim}
\mathbf{J} - \mathbf{N}\mathbf{K}
\end{equation}
where $G_{aa^{\prime}}(R)$ is the matrix of solutions formed from $G_{a}(R)$ with the 
second subscript $a^{\prime}$ labelling the linearly independent solutions generated 
by different choices of boundary conditions. The real diagonal matrices $\mathbf{J}$
and $\mathbf{N}$ are given by
\begin{equation}
  \begin{aligned}
  J_{aa} &=  \mathcal{C}_\Gamma \kappa_\Gamma^{1/2} R j_{l}(k_\Gamma R), \\
  N_{aa} &=  \mathcal{D}_\Gamma \kappa_\Gamma^{1/2} R n_{l}(k_\Gamma R),
  \end{aligned}
\end{equation}
where $\kappa_\Gamma = |k_\Gamma|$ and $j_l(z)$ and $n_l(z)$ are the regular and
irregular spherical Bessel functions. For open channels ($k_\Gamma^2 > 0$), the
Bessel functions are oscillatory and $\mathcal{C}_a = \mathcal{D}_a = 1$ and for
closed channels ($k_\Gamma^2 < 0$) they are exponentially increasing and
decreasing functions with $\mathcal{C}_\Gamma = (-i)^l$ and
$\mathcal{D}_\Gamma = i^{l+1}$. The reactance matrix $\mathbf{K}$ is of
dimension $N_{T}\times N_{T}$ where $N_{T}=N_{o}+N_{c}$ is the total number of
open $N_{o}$ and closed $N_{c}$ channels.

The required open channel scattering matrix
$\mathbf{S}_{oo}$ is obtained from the reactance matrix
$\mathbf{K}$ by \cite{Mies1980}
\begin{equation}
\label{R35}
\mathbf{S}_{oo} = (\mathbf{I}+i [\mathbf{K}_{oo}+\mathbf{K}_{oo}^{R}])
(\mathbf{I}-i [\mathbf{K}_{oo}+\mathbf{K}_{oo}^{R}])^{-1}  
\end{equation}
where 
\begin{equation}
\label{R35a}
\mathbf{K}_{oo}^{R} = -\mathbf{K}_{oc}[\mathbf{I}+\mathbf{K}_{cc}]^{-1} \mathbf{K}_{co}
\end{equation}
embodies the effects of the closed channels on $\mathbf{S}_{oo}$. The asymptotic
fitting for the open channels requires very large values of $R$ where the closed
channel contributions must be absent. Thus,
$\mathbf{K}_{cc} \approx \mathbf{I}$, $\mathbf{K}_{oc} \approx \mathbf{0}$ and
$\mathbf{K}_{oo}^{R} \approx \mathbf{0}$ so that only the open channel components of
$\mathbf{K}$ are needed for the determination of $\mathbf{S}_{oo}$.

For systems formed from either ${}^{4}$He$^{*}$, or ${}^{3}$He$^{*}$ trapped in its 
lower hyperfine level with energy $E > V_\Gamma^\infty$, the scattering channels will be open. 
However, with our choice of energy origin, closed channels occur at low energies 
for the 3--3 and 3--4 systems. The closed channels must be included in the multichannel 
equations as this coupling may be quite significant at smaller values of $R$.
 
As a result of the complex optical potential, both $\mathbf{G}$ and $\mathbf{S}$ 
are complex. However, note that $\mathbf{S}$ remains symmetric.

\section{Perturbed single channel model}
\label{sec:single-channel}

We now consider a perturbed single channel model~\cite{Peach2017} for 
$\mathrm{He}^{*} + \mathrm{He}^{*}$ scattering in the states
${}^{1}\Sigma_{g}$, ${}^{3}\Sigma_{u}$, ${}^{5}\Sigma_{g}$.  The hyperfine 
couplings and splittings due to the ${}^{3}$He$^{*}$ nuclear spin are neglected 
but the constraints due to the different quantum statistical symmetries are 
included.
If the spin dipole interaction is ignored,  the radial 
functions $F_{kl}^{S}(R)$ for the scattering satisfy 
\begin{equation}
\label{P1}
\left[ \frac{d^{2}}{dR^{2}} - \frac{l(l+1)}{R^{2}} - \frac{2 \mu \;{}^{2S+1}V_{\Sigma}(R)}
{\hbar^{2}} + k^{2} \right] F_{kl}^{S}(R) =0,
\end{equation}
where $S=0,1,2$ and $k=\sqrt{2 \mu E}/\hbar$. 
At large $R$, the radial functions have the asymptotic form (c.f. (\ref{R34}))
\begin{equation}
\label{P2}
F_{kl}^{S}(R) \underset{R\rightarrow\infty}{\sim} \frac{1}{\sqrt{k}} 
\left[ (kR) j_{l}(kR) \cos \eta_{l}^{S} - (kR) n_{l}(kR) \sin \eta_{l}^{S} \right].
\end{equation}
The phase shifts $\eta_{l}^{S}(k)$ are complex for $S=0,1$ since ${}^{1,3}V_{\Sigma}(R)$ are 
complex, whereas, for $S=2$, ${}^{5}V_{\Sigma}(R)$ is real and so  $\eta_{l}^{2}(k)$ is
also real. The single (open) channel scattering matrix is given by
\begin{equation}
\label{P3a} 
\mathbf{S} = \exp{[2i \eta_{l}^{S}(k)}], 
\end{equation}
where $|\mathbf{S}|<1$ for $S=0,1$ and $|\mathbf{S}|=1$ for $S=2$.

If we introduce the perturbation produced by the spin-dipole interaction, the channels
with $S=0,2$ become coupled, whereas the states with $S=1$ give rise to a separate $S$-matrix
in which the spin-dipole interaction is retained although in this case it does not affect 
the final collision rate very much. The scattering matrices then have the form~\cite{Seaton1966}
\begin{equation}
\label{P3}
\mathbf{S}=e^{i \mathbf{\eta}}(\mathbf{I}+i\mathbf{K}^{\mathrm{sd}})
(\mathbf{I}-i\mathbf{K}^{\mathrm{sd}})^{-1}e^{i\mathbf{\eta}},
\end{equation}
where $\exp{(i\mathbf{\eta})} $ is a diagonal matrix with elements $\eta^{S}_{l}(k)$. 
So far no approximation has been made in writing $\mathbf{S}$ in this form.

We now use Born perturbation theory to approximate $\mathbf{K}^{\mathrm{sd}}$ 
by~\cite{Seaton1961} 
\begin{equation}
\label{P4}
K^{\mathrm{sd}}_{\tilde{a}^{\prime},\tilde{a}}(k) = \frac{2 \mu}{\hbar^{2}}  
\langle \psi_{k,\tilde{a}^{\prime}}| \hat{H}_{\text{sd}} | \psi_{k,\tilde{a} }
\rangle
\end{equation}
where $\tilde{a}=\{\gamma,S,l,J,M_{J}\}$ and the unperturbed state eigenfunction is 
\begin{equation}
\label{P5}
|\psi_{k,\tilde{a}}(R) \rangle = R^{-1} F_{kl}^{S}(R) |\tilde{a} \rangle .
\end{equation}
Using (\ref{R26a}) then gives
\begin{eqnarray}
\label{P6}
K^{\mathrm{sd}}_{\tilde{a}^{\prime},\tilde{a}}(k) & = & -\left(\frac{2\mu }{\hbar^{2}}\right)  
D_{\tilde{a}^{\prime}\tilde{a} }  \nonumber  \\
&& \times \int_{0}^{\infty} dR \;F_{kl^{\prime}}^{S^{\prime}}(R)^{*}V_{p}(R) 
F_{kl}^{S}(R).
\end{eqnarray}
where $D_{\tilde{a}^{\prime}\tilde{a} }$ is given by (\ref{R26d}). In the evaluation 
of the radial integral any non-zero contribution from the imaginary parts of the 
radial functions $F_{kl^{\prime}}^{S^{\prime}}(R)$ and $F_{kl}^{S}(R)$ is neglected.
This makes only a very small difference to the result for 
$K^{\mathrm{sd}}_{\tilde{a}^{\prime},\tilde{a}}(k)$. The $S$-matrix (\ref{P3})
is then evaluated without further approximation. 

The elastic and ionization cross sections are given by (\ref{R31}) 
and (\ref{R32}) respectively, with $f=S$ and $f^{\prime}=S$.

So far we have not considered the application of this theory to the different 
isotopic combinations. These differences introduce changes to the interpretation
of the sums over $l$ and $l^{\prime}$ where the spin-dipole interaction imposes the 
condition $|l-l^{\prime}|=0,2$. This interpretation depends upon the behavior of the system 
under $\hat{X}_{N}$ which permutes the nuclear labels, interchanging the nuclear spins
and reversing the molecular axis~\cite{Geltman1969}.
For the bosonic 4--4 system the wavefunction must be symmetric under $\hat{X}_{N}$ and,
as there is no nuclear spin, $l$ must be even (odd) for $S$ even (odd). Thus, for 
the ${}^{1}\Sigma_{g}$ and ${}^{5}\Sigma_{g}$ states,
\begin{equation}
\label{P7}
\sum_{(l)}= 2 \sum_{l \,\text{even}},  
\end{equation}
whereas, for the ${}^{3}\Sigma_{u}$ state,
\begin{equation}
\label{P8}
\sum_{(l)}= 2 \sum_{l \,\text{odd}}.  
\end{equation}
For the fermionic 3--3 system, the wavefunction must be antisymmetric and, as the 
total nuclear spin forms antisymmetric singlet (symmetric triplet) states for $i=0 (1)$, the sum is
\begin{equation}
\label{P9}
\sum_{(l)} = \frac{1}{2}\sum_{l\,\text{even}} + \frac{3}{2} \sum_{l \,\text{odd}}
\end{equation}
for ${}^{1}\Sigma_{g}$ and ${}^{5}\Sigma_{g}$ states, and
\begin{equation}
\label{P10}
\sum_{(l)} = \frac{1}{2}\sum_{l\,\text{odd}} + \frac{3}{2} \sum_{l \,\text{even}}
\end{equation}
for the ${}^{3}\Sigma_{u}$ state.
For the heteronuclear 3--4 system there is no symmetry under $\hat{X}_{N}$ and
$\sum_{(l)}$ is to be interpreted for ${}^{1}\Sigma_{g}$, ${}^{3}\Sigma_{u}$, 
and ${}^{5}\Sigma_{g}$ as a sum over all $l$.  

\section{Results and Discussion}
\label{sec:results}

\subsection{Cross sections}

Details of the integration of the coupled multichannel equations (\ref{R10}) are
given in Appendix B. The calculations require as input the BO molecular
potentials ${}^{1,5}\Sigma_{g}^{+}$ and ${}^{3}\Sigma_{u}^{+}$, and total ionization
widths ${}^{1,3}\Gamma (R)$. 

The molecular potential for ${}^5 V_{\Sigma}(R)$ is taken from the accurate
calculations of Przybytek and Jeziorski~\cite{Przy2005}, which include adiabatic
and relativistic corrections. The ${}^{1,3} V_{\Sigma}(R)$ potentials were
constructed by taking the tabulated potentials of M\"uller \textit{et
  al.}~\cite{Muller1991}, available only for the short-range region
$R < 14$~$a_0$, and matching them onto the long-range form of the
${}^5 V_{\Sigma}(R)$ potential. An exchange term of the form
${}^{1,3}V_{\text{exch}}(R)=A_{1,3}\exp{(-\beta_{1,3} R)}$ was included such
that the potentials have the form
${}^{1,3}V_{\Sigma}(R>14\,a_0) = {}^5 V_{\Sigma}(R) -
{}^{1,3}V_{\text{exch}}(R)$. After fitting to the last two points of the
tabulated data, the exchange coefficients were found to be $A_1 = 5.9784$,
$\beta_1 = 0.7367$, $A_3 = 1.7980$ and $\beta_3 = 0.6578$.

\begin{figure}[ht]
\includegraphics[width=0.95\columnwidth]{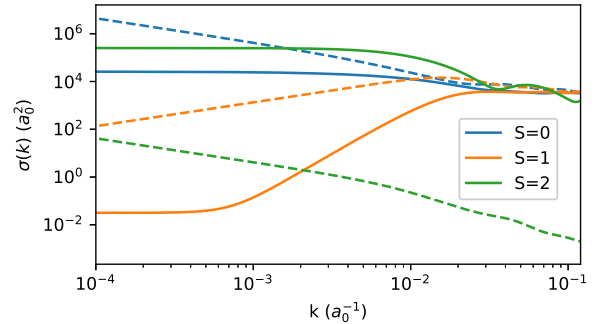}
\caption{\label{fig:xsect_44}Momentum dependence of elastic cross sections
  (solid lines) given by
  (\ref{R31}) and ionization cross sections (dashed lines) given by (\ref{R32}) for the
  4--4 system.}
\end{figure}
\begin{figure}[ht]
\includegraphics[width=0.95\columnwidth]{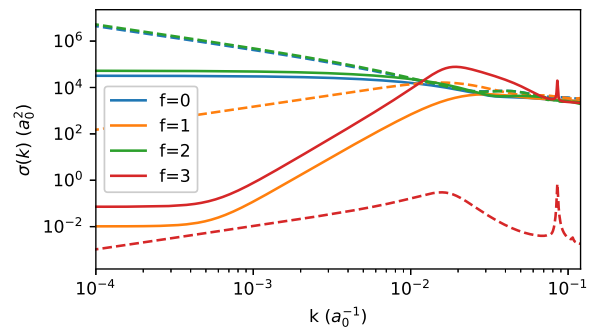}
\caption{\label{fig:xsect_33}Cross sections for the 3--3 system, as in figure~\ref{fig:xsect_44}, with both
 ${}^{3}$He$^{*}$ atoms in their lower hyperfine level. Note the sharp
 structures at around ~$10^{-1}$~$a_0$ are a result of resonances in the
 potentials for scattering with $l=5$.}
\end{figure}
\begin{figure}[ht]
\includegraphics[width=0.95\columnwidth]{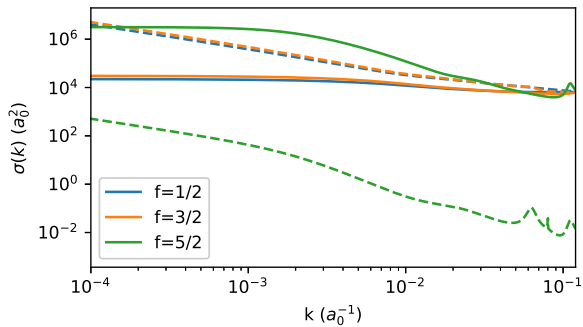}
\caption{\label{fig:xsect_34}Cross sections for the 3--4 system, as in figure~\ref{fig:xsect_33}, with the
 ${}^{3}$He$^{*}$ atom in its lower hyperfine level.}
\end{figure}

The calculated $T$- and $S$-matrix elements were used to determine cross
sections for scattering. For the 4--4 system we have
$\Gamma =\{S_{1},S_{2}\} = (1,1)$ and $f=S=0,1,2$; for the 3--3 system trapped in
the lower hyperfine level,
$\Gamma =\{S_{1},i_{1},f_{1},S_{2},i_{2},f_{2}\}=(1,\frac{1}{2},\frac{3}{2},1,
\frac{1}{2},\frac{3}{2})$ and $f=0,1,2,3$; for the 3--4 system trapped in the
lower hyperfine level,
$\Gamma=\{S_{1},i_{1},f_{1},S_{2}\} = (1,\frac{1}{2},\frac{3}{2},1)$ and
$f=\frac{1}{2},\frac{3}{2},\frac{5}{2}$. Cross sections calculated from
(\ref{R31}) for elastic scattering and (\ref{R32}) for ionization of the 4--4,
3--3 and 3--4 systems in the lowest hyperfine level are shown in Figs
\ref{fig:xsect_44}, \ref{fig:xsect_33} and \ref{fig:xsect_34} respectively.

At low energies the $T$-matrix elements used in the calculation of the rates should
have an energy dependence determined by the Wigner threshold behavior of the
$T$-matrix elements for a $R^{-n}$ interaction, that is~\cite{Mies2000}
\begin{equation}
\label{P20}
|T_{aa}|^{2} \underset{k_{a}\rightarrow 0}{\sim}  \left\{
			\begin{array}{lll}
			\left[k_{a}^{2l+1}\right]^{2}, & (2l+1) < (n-2), & n > 3  \\
			\left[k_{a}^{n-2}\right]^{2}, & (2l+1) > (n-2),  & n > 3  \\
			\left[k_{a} \ln k_{a}\right]^{2}, & l=0, & n=3 \\
			k_{a}^{2},  & l \geq 1,& n=3
			\end{array}
			\right .
\end{equation}
and
\begin{equation}
\label{P21}
	|T_{a^{\prime} a}|^{2} \underset{k_{a}\rightarrow 0}{\sim} 
	k_{a}^{2l+1}, \quad a^{\prime} \neq a .
\end{equation}

In all cases, a dependence of $k^0_\Gamma$ is observed as
  $k_\Gamma \rightarrow 0$ in the elastic cross sections, however this is a
  result of different processes. The behavior of the cross sections for the
homonuclear systems is a consequence of the selection rules $(-1)^{l-S}=1$ for
the 4--4 system and $(-1)^{l-f}=-1$ for the 3--3 system (with
$f_{1}=f_{2}=\frac{3}{2}$), which require $l$ even (odd) for $S$ even (odd) for
the 4--4 system and $l$ odd (even) for $f$ even (odd) for the 3--3 system.
Since the matrix elements (\ref{R23}) of the spin-dipole $R^{-3}$ interaction
vanish for $s$-wave elastic scattering but are non-zero for elastic $p$-wave
scattering, the threshold behavior of the elastic cross sections is only
determined by the $R^{-3}$ interaction when $s$-wave scattering is excluded (as
is the case for the 4--4 system with $S=1$ and the 3--3 system with $f=0,2$), in
which case the dependence is given by $k_{\Gamma}^{0}$.  When $s$-wave
scattering is present (the 4--4 system with $S=0,2$, the 3--3 system with
$f=1,3$, and the 3--4 system where there is no selection rule) the threshold
behavior is due to the long-range $R^{-6}$ interaction and the elastic cross
sections have the variation $k_{\Gamma}^{0}$.

At higher energies where elastic $p$-wave scattering is due to the $R^{-6}$
interaction, the elastic cross sections have a $k_{\Gamma}^{4}$ dependence. As
the inelastic (ionization) cross sections have the threshold behavior
$k_{\Gamma}^{2l+1}$~\cite{Julienne1989,Mies2000}, the $s$-wave ionization cross
sections vary as $k_{\Gamma}^{-1}$ at very low energies whereas the $p$-wave
cross sections vary as $k_{\Gamma}^{1}$.

There are also several peaks observable near $k=0.1~a_0^{-1}$. We have
identified these as resonances that occur in the $l=5$ partial wave for the
${}^5 \Sigma_g$ and ${}^3 \Sigma_u$ potentials. We note the selection rule in
the 4--4 system suppresses this resonance in the $S=2$ case. 

\subsection{Rates}

\begin{figure}[ht]
\includegraphics[width=0.95\columnwidth]{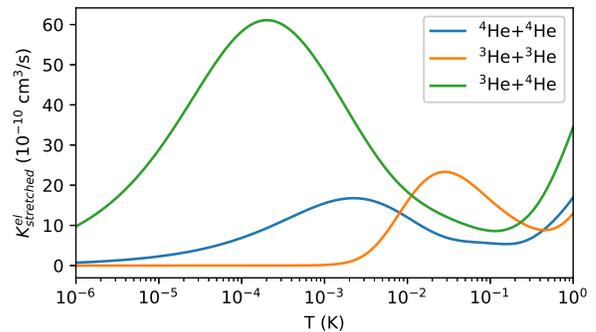}
\caption{\label{fig:ss_el}Thermally averaged spin-stretched elastic rates for the 3--3, 3--4 and 4--4 systems.}
\end{figure}

\begin{figure}[ht]
\includegraphics[width=0.95\columnwidth]{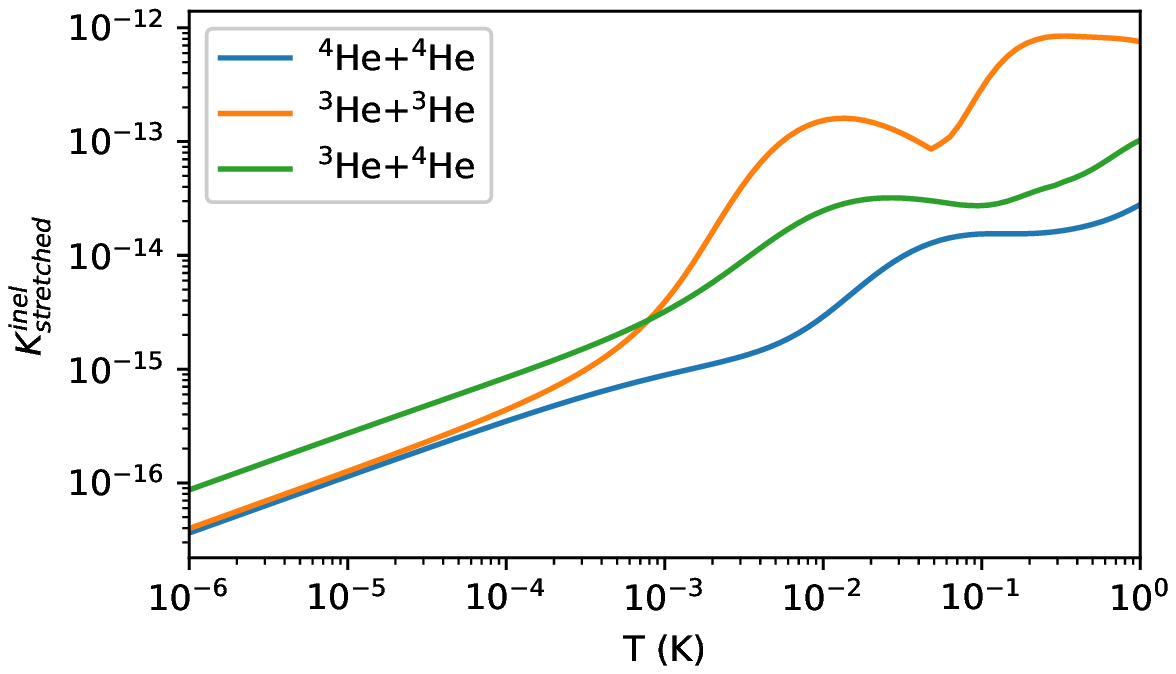}
\caption{\label{fig:ss_inel}Thermally averaged spin-stretched inelastic rates for the 3--3, 3--4 and 4--4 systems.}
\end{figure}

\begin{figure}[ht]
\includegraphics[width=0.95\columnwidth]{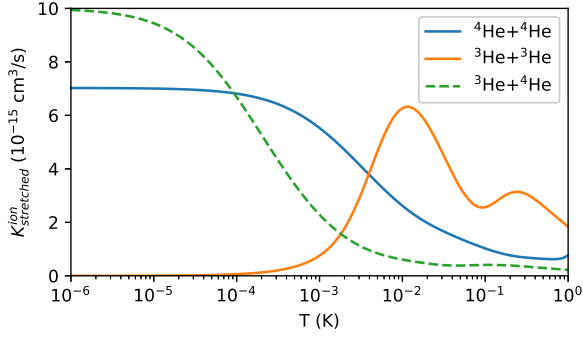}
\caption{\label{fig:ss_lo}Thermally averaged spin-stretched ionization rates for the 3--3, 3--4 and 4--4 systems. 
The dashed line for the 3--4 rate indicates the rate is a factor of 10 larger than shown in the graph.}
\end{figure}

\begin{figure}[ht]
\includegraphics[width=0.95\columnwidth]{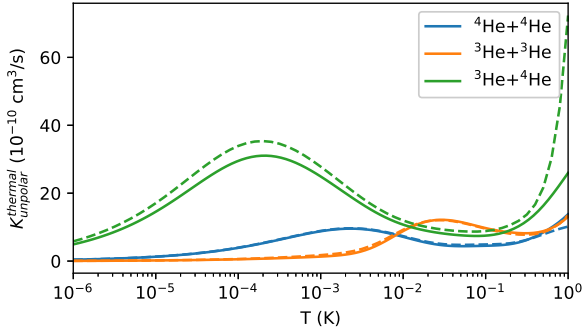}
\caption{\label{fig:un_thermal}Thermally averaged total unpolarized thermalization
  rates for the 3--3, 3--4 and 4--4 systems. Solid lines correspond to the
  multichannel calculation and dashed lines to the single-channel calculation.}
\end{figure}

\begin{figure}[ht]
\includegraphics[width=0.95\columnwidth]{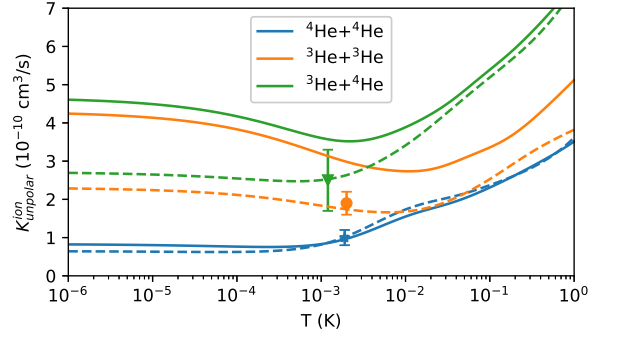}
\caption{\label{fig:un_lo}Thermally averaged total unpolarized ionization rates
  for the 3--3, 3--4 and 4--4 systems.  Solid and dashed lines correspond to the
  multichannel and single-channel calculations respectively. Experimental data
  points are shown for the 4--4 (star, \cite{Stas2006}), 3--3 (circle,
  \cite{Stas2006}) and 3--4 (triangle, \cite{McNamara2007}) systems. Note that
  these experimental data points were extrapolated from a mixed population of
  magnetic sublevels to the completely unpolarized mixture using an approximate
  theory. Our unpolarized rates should be viewed as an upper bound for any
  mixture of magnetic sublevels.}
\end{figure}

In Figures.~\ref{fig:ss_el}--\ref{fig:ss_lo} we report thermally averaged
elastic, inelastic and ionization rates respectively for spin-stretched initial
states, that is, $S=2$ for the 4--4 system, $f=3\,(S=2)$ for the 3--3 system,
and $f=\frac{5}{2}\,(S=2)$ for the 3--4 system. It can be seen that the
ionization rates are much smaller than the elastic rates. In addition, the
inelastic rates, which indirectly contribute to ionization, are even smaller for
low temperatures. At higher temperatures these inelastic rates are more
important, representing the dominant pathway for ionization, although they
remain smaller than the elastic rates. The relative magnitudes of these cross
sections are the reason for long lifetimes of a spin-stretched gas of metastable
helium.

The unpolarized thermal and ionization rates are shown in
Fig.~\ref{fig:un_thermal} and Fig.~\ref{fig:un_lo} respectively.  As expected,
the ionization rates for polarized systems in a spin-stretched initial state are
strongly suppressed compared to those for unpolarized systems, the suppression
being $O(10^{-4})$ for the 4--4 and 3--4 systems and $O(10^{-5})$ for the 3--3
system.

The single-channel calculations are shown in Fig.~\ref{fig:un_thermal} and
Fig.~\ref{fig:un_lo} as dashed lines. We can see reasonable agreement with the
4--4 system, in which there is little difference between the couplings included
in the multichannel and single-channel formalism, but there are much larger
disagreements for the 3--3 and 3--4 systems. We believe this originates in the
effective diabatic connection between the outer and inner regions of the
calculation, which will be discussed in more detail in
section~\ref{sec:othercalcs}.

\begin{figure}[ht]
\includegraphics[width=0.95\columnwidth]{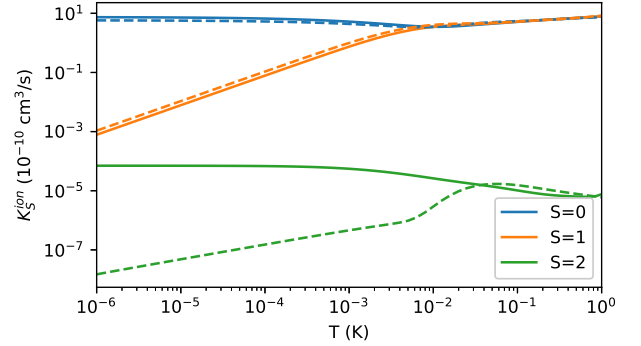}
\caption{\label{fig:sc-comp-44}The multichannel and single-channel ionization
  rates for the 4--4 system shown as solid and dashed lines, respectively. Here
  $K^\textrm{loss}_S(T) = K(T;\Gamma,S,S \rightarrow \mathrm{PI})$. The
  disagreement for the $S=2$ channel is due to the perturbative treatment of the
  spin-dipole coupling.}
\end{figure}
We also show a comparison of unpolarized ionization rates for the 4--4 system
between the single-channel and multichannel calculations in
Fig.~\ref{fig:sc-comp-44}. We can see that the single-channel calculation
performs well for the $S=0,1$ channels, but fails to capture the correct
$T\rightarrow 0$ form of the $S=2$ channel. This is due to the perturbative
treatment of the spin-dipole coupling, connecting the $S=2$ channel to the
$S=0$. We believe the radial-dependence of the ionization process is not well
captured in the perturbative treatment as the scattering wave functions
  in the $l=2$ states do not represent very well the short-range properties of
  the complex singlet potential.

Our values for rates with spin-stretched and unpolarized mixtures should place rough
lower and upper bounds, respectively, on the rates for an arbitrary mixture of
magnetic sublevels in an experimental configuration.

\subsection{Comparison with other calculations}
\label{sec:othercalcs}

The present calculations for the 4--4 system essentially reproduce the results
of Venturi~\etal~\cite{Venturi2000}
although there are some differences arising from our use of the Przybytek and 
Jeziorski~\cite{Przy2005}  ${}^5 V_{\Sigma}(R)$ potential rather than the older St\"{a}rck 
and Meyer~\cite{SM1994} potential used by~\cite{Venturi2000}.

Our calculated total unpolarized ionization coefficients are higher than those
calculated by~\cite{Stas2006,McNamara2007,Dickinson2007} using a two-stage
semiclassical model, the differences at $T=1$~mK, for example, being
approximately 30\% and 15\% larger for 3--3 system and 3--4 system respectively.
As the semiclassical models assume 100\% ionization at small $R$ in the $S=0,1$
states, in contrast to the use of a complex optical potential, it has been
argued~\cite{Dickinson2007} that these semiclassical models should give upper
bounds to the ionization coefficients. However, we note that the calculations of
\cite{Stas2006,McNamara2007,Dickinson2007} answer the question ``what proportion
of incoming flux will pass through to the short-range in the singlet or triplet
state'' by assuming a diabatic connection between the outer basis, best
described by $f_1$ and $f_2$, and the inner basis, best described by $S$ and
$i$. This is effectively a two-stage or ``single-pass'' model as all flux in the
$S=0,1$ states at short range is completely lost through ionization while all
flux in the $S=2$ state is assumed to be reflected outwards and leave the
scattering region. Our single-channel calculations, through the sum
  over $l$ in equations \eqref{P7}--\eqref{P10}, effectively apply this same
  description of a diabatic connection from the outer to inner basis.

We argue that the question ``what proportion of incoming flux can return as
outgoing flux'' should instead be considered. This requires a three-stage model
with a second diabatic connection from the inner to outer basis, where some
states (i.e. higher hyperfine levels) are energetically forbidden. In these
channels, the outgoing flux would be reflected inwards, remaining trapped in the
system for multiple ionization attempts. Hence, some of our multichannel values
are larger than those predicted by the semiclassical models.

\subsection{Comparison with experimental measurements}

The elastic collision rate for the spin-polarized 4--4 system at
$T=1.0 \pm 0.1 $ mK has been measured \cite{Browaeys2001} to be
$ \alpha = 5 \times 10^{-9}$ cm$^{3}$/s to within a factor of 3.  This compares
with our theoretical value of $\alpha = 2K = 1.76 \times 10^{-9}$
cm$^{3}$/s. This is just within the range of experiment, although we note that
small variations to the short-range parts of the potential
\cite{Venturi2000,Cocks2010} can affect this.

A comparison of our calculated loss rates with the various calculated and measured loss rate 
coefficients reported in the literature is given in Table~\ref{tab:Kvalues}. 
\begin{table}
  \caption{\label{tab:Kvalues}Thermally averaged total unpolarized ionization
    rates for the 3--3, 3--4 and 4--4 systems in units of $10^{-10}$
    cm$^{3}$/s. Results from our present calculations, using ionization widths
    $\Gamma_{\text{M}}(R)$ and $\Gamma_{\text{GMS}}(R)$, denoted
    $K^{\text{th}}_{\text{M}}$ and $K^{\text{th}}_{\text{GMS}}$ respectively,
    and single-channel calculations $K^{\text{th}}_{\text{SC}}$ are compared to
    experimental values $K^{\text{exp}}$ and calculated values using simple
    two-stage semiclassical models
    $K^{\text{th}}_{\text{2S}}$~\cite{KM1999,Stas2006,McNamara2007,Mastwijk1998,Tol1999}.  Numbers in
    brackets denote one standard deviation experimental errors.}
\begin{ruledtabular}
\begin{tabular}{llZZNNZZ}
&
Ref. &
\multicolumn{1}{c}{\text{T(mK)}} &
\multicolumn{1}{c}{$K^{\text{exp}}$} &
\multicolumn{1}{c}{$K^{\text{th}}_{\text{M}}$} &
\multicolumn{1}{c}{$K^\text{th}_\text{GMS}$} &
\multicolumn{1}{c}{$K^{\text{th}}_{\text{SC}}$} &
\multicolumn{1}{c}{$K^{\text{th}}_{\text{2S}}$}
  \\
\hline
4--4 & \cite{Mastwijk1998} & 1.0    & 2.7(1.2) & 0.831 & 0.830 & 0.82 &      \\
     & \cite{Tol1999}      & 1.0    & 1.3(0.2) & 0.831 & 0.830 & 0.82 &      \\
     & \cite{KM1999}       & 0.5    & 3.8(1.1) & 0.769 & 0.792 & 0.70 & 2.2  \\
     & \cite{Stas2006}     & 1.9(1) & 1.0(2)   & 0.931 & 0.956 & 1.01 & 0.80 \\
3--3 & \cite{KM1999}       & 0.5    & 11.0(4)  & 3.23  & 3.41  & 1.94 & 11   \\
     & \cite{Stas2006}     & 2.0(3) & 1.9(3)   & 2.90  & 2.98  & 1.74 & 1.8  \\
3--4 & \cite{McNamara2007} & 1.2(1) & 2.9(8)   & 3.61  & 3.57  & 2.52 & 2.9  \\
\end{tabular}
\end{ruledtabular}
\end{table}

The present loss rate coefficients are in good agreement with the measurements
of~\cite{Stas2006} for the 4--4 system, but our 3--4 and 3--3 results are above the
measurements of \cite{McNamara2007} and~\cite{Stas2006}. However, we note that
these experiments were carried out in a MOT where the magnetic sublevel mixture
was not an unpolarized set. These papers used the semiclassical theory discussed
above to then rescale their experimental results to estimate an experimental
unpolarized rate, however we believe this has underestimated the true value.
Note that there are significant discrepancies between the experimental values,
due possibly to approximations in the experimental analysis such as the neglect
of the magnetic substate distribution~\cite{Stas2006}. The large discrepancy
with the 3--3 and 4--4 semiclassical calculations of~\cite{KM1999} is not
surprising as these calculations are based upon several incorrect assumptions
that significantly overestimate the rate coefficients~\cite{Stas2006}.

\section{Summary}
\label{sec:summary}

Scattering and ionizing cross sections and rates have been calculated for ultracold 
collisions between metastable helium atoms using a fully quantum-mechanical 
close-coupled formalism. Homonuclear collisions of the bosonic 
${}^{4}$He$^{*} +{}^{4}$He$^{*}$ and fermionic 
${}^{3}$He$^{*} + {}^{3}$He$^{*}$ systems, and heteronuclear collisions of the mixed 
${}^{3}$He$^{*} +{}^{4}$He$^{*}$ system, were investigated over a temperature range 1~$\mu$K to 1~K. 
Carefully constructed Born-Oppenheimer molecular potentials were used to describe the electrostatic 
interaction between the colliding atoms. The loss through ionization from the 
${}^{1,3}\Sigma $ states was represented by complex optical potentials. 
Magnetic spin-dipole mediated  transitions from the ${}^{5}\Sigma $ state were included and 
results obtained for spin-polarized and non spin-polarized systems.

The calculated scattering and ionization cross sections have the appropriate Wigner  threshold behavior
for momenta below $k \approx 10^{-3}\,a_{0}^{-1}$, and exhibit several peaks near $k=0.1\;a_{0}^{-1}$,
identified as resonances in the $l=5$ partial wave.

Thermally averaged rates for spin-stretched initial states ($S=2$ for the 4--4 system, 
$f=3$ for the 3--3 system, and $f=\frac{5}{2}$ for the 3--4 system) are greatest for elastic scattering,
$O(10^{-5})$ smaller for ionization, and range from $O(10^{-7})$ smaller at 1 $\mu$K to
$O(10^{-5})$ smaller at 1 K for inelastic scattering. We note that there is a
significantly larger ionization rate of the 3--4 system, which leads to
stronger losses for dual-species mixtures. The thermally averaged rates for
unpolarized systems are enhanced by $O(10^{4})$ for the 4--4 and 3--4 systems, and $O(10^{5})$
for the 3--3 system, compared to the spin-stretched rates.

The total unpolarized ionization rates are higher than those calculated using two-stage semiclassical
models~\cite{Stas2006,McNamara2007} based upon a diabatic connection between the basis states
in the inner and outer regions. It has been argued~\cite{Dickinson2007} that these semiclassical 
models should give upper bounds on ionization rates but we suggest that a three-stage semiclassical 
model which includes a second diabatic connection is more appropriate and that such a model 
would give higher rates.

Finally, a perturbed single channel model was developed in which hyperfine
couplings and splittings are neglected but the effects of the different quantum
statistical symmetries are included. It was found that this single-channel
approximation follows a similar trend to the two-stage semi-classical model and
underestimates the ionization rates for the 3--3 and 3--4 systems.

\appendix

\section{Basis states and matrix elements}

Matrix elements of $\hat{H}_\text{el}$ and $\hat{H}_{\text{sd}}$ in the $\{|a_{12}\rangle\}$
basis (\ref{R16a}) are required whereas they are most easily evaluated in the 
$\{|\alpha \rangle \}$ basis (\ref{R22}). The first basis 
\begin{equation}
\label{A1}
|a_{12}\rangle = |(S_{1},i_{1},f_{1})_A;(S_{2},i_{2},f_{2})_B;f,l,J,M_J\rangle 
\end{equation}
involves the couplings
\begin{equation}
\label{A2}
\hat{\mathbf{S}}_{1}+\hat{\mathbf{i}}_{1}=\hat{\mathbf{f}}_{1}, \quad 
\hat{\mathbf{S}}_{2}+\hat{\mathbf{i}}_{2}=\hat{\mathbf{f}}_{2}, \quad 
\hat{\mathbf{f}}_{1}+\hat{\mathbf{f}}_{2}=\hat{\mathbf{f}}, 
\end{equation}
with associated states
\begin{equation}
\label{A2a}
|(S_{1},i_{1},f_{1})_A;(S_{2},i_{2},f_{2})_B;f,m_{f}\rangle  ,
\end{equation}
whereas the second basis 
\begin{equation}
\label{A3} 
|\alpha \rangle = |(S_{1})_A,(S_{2})_B; S, M_{S} \rangle |l,m_{l}\rangle
\end{equation}
is associated with the couplings
\begin{equation}
\label{A3a}
\hat{\mathbf{S}}_{1}+\hat{\mathbf{S}}_{2}=\hat{\mathbf{S}},  \quad
\hat{\mathbf{i}}_{1}+\hat{\mathbf{i}}_{2}=\hat{\mathbf{i}}, \quad
\hat{\mathbf{S}}+\hat{\mathbf{i}}=\hat{\mathbf{f}}
\end{equation}
and the states
\begin{widetext}
\begin{equation}
\label{A6}
|(S_{1},i_1)_A;(S_{2},i_{2})_B;S,i;f,m_{f}\rangle  = \sum_{M_{S},m_{i}}C^{Sif}_{M_S,m_i,m_f} 
|(S_{1})_A,(S_{2})_B; S, M_{S} \rangle |(i_{1})_{A},(i_{2})_{B};i,m_{i} \rangle .
\end{equation}
The relationship between states in the two couplings (\ref{A2}) and (\ref{A3a}) is~\cite{Brink68}
\begin{equation}
\label{A5}
|(S_{1},i_{1},f_{1})_A;(S_{2},i_{2},f_{2})_A;f,m_{f}\rangle  =  \sum_{S,i} [Sif_{1}f_{2}]^{1/2} 
\Ninej{S_1}{S_2}{S}{i_1}{i_2}{i}{f_1}{f_2}{f} 
|(S_{1},i_1)_A;(S_{2},i_{2})_B;S,i;f,m_{f}\rangle .
\end{equation}
Finally, the first basis requires the coupling $\hat{\mathbf{f}}+\hat{\mathbf{l}}=\hat{\mathbf{J}}$
to give
\begin{equation}
\label{A4}
|(S_{1},i_{1},f_{1})_A;(S_{2},i_{2},f_{2})_B;f,l,J,M_J\rangle =\sum_{m_{f},m_{l}}
C^{flJ}_{m_f,m_l,m_J} 
 |(S_{1},i_{1},f_{1})_A;(S_{2},i_{2},f_{2})_B;f,m_{f}\rangle |l,m_{l}\rangle .
\end{equation}
Hence
\begin{eqnarray}
\label{A7}
\langle a_{12}^\prime | \hat{H}_\text{el} | a_{12} \rangle & = & 
\delta_{i_{1}^{\prime},i_{1}} \delta_{i_{2}^{\prime},i_{2}} \delta_{l^{\prime},l}
			[f_1^\prime f_2^\prime f_1 f_2]^{1/2} \; \sum_{S^{\prime},S,i}[S^{\prime}S]^{1/2}[i] 
			\nonumber  \\
			&& \times \sum_{m_{f}^{\prime},m_f,m_{l}} C^{f^\prime l J^{\prime}}_{m_f^{\prime},m_l,m_J^{\prime}} 
			C^{flJ}_{m_f,m_l,m_J}  \sum_{M_{S}^{\prime},M_{S},m_{i}} 
			C^{S^{\prime}if^{\prime}}_{M_{S}^{\prime},m_i,m_f^{\prime}} C^{Sif}_{M_S,m_i,m_f} 
			\nonumber  \\
			 && \times \Ninej{S_1}{S_2}{S}{i_1}{i_2}{i}{f_1^\prime}{f_2^\prime}{f^{\prime}}
									\Ninej{S_1}{S_2}{S}{i_1}{i_2}{i}{f_1}{f_2}{f} 
									\;\langle (S_{1}^{\prime})_A,(S_{2}^{\prime})_B,S^{\prime},	M_{S}^{\prime}|	
									\hat{H}_\text{el} |(S_{1})_A,(S_{2})_B;S,M_{S} \rangle	.		
\end{eqnarray}
Transforming the matrix element into body-fixed states using (\ref{R18a})	and then
using their eigenvalue equation (\ref{R18}) gives
\begin{equation}
\label{A7a}
\langle (S_{1}^{\prime})_A,(S_{2}^{\prime})_B,S^{\prime},	M_{S}^{\prime}|	\hat{H}_\text{el} |(S_{1})_A,(S_{2})_B;S,M_{S} \rangle										
		= \delta_{S_{1}^{\prime},S_{1}} \delta_{S_{2}^{\prime},S_{2}} \delta_{S^{\prime},S}
		\sum_{\Omega_{S}} D^{S\,\dagger}_{M_{S}^{\prime},\Omega_{S}} (\phi,\theta,0) 
		D^{S}_{\Omega_{S},M_{S}} (\phi,\theta,0) \,{}^{2S+1}V_\Sigma (R).
\end{equation}
The unitarity of the rotation matrix gives $\delta_{M_{S}^{\prime},M_{S}}$, the summations over the 
Clebsch-Gordan coefficients $\delta_{f^{\prime},f}\,\delta_{J^{\prime},J}\, \delta_{M_{J}^{\prime},M_{J}}$,
and (\ref{A7}) reduces to (\ref{R18b}).

Similarly, the matrix element of $\hat{H}_{\text{sd}}$ is
\begin{eqnarray}
\label{A8}
\langle a_{12}^\prime | \hat{H}_\text{sd} | a_{12} \rangle & = & \delta_{i_{1}^{\prime},i_{1}} 
\delta_{i_{2}^{\prime},i_{2}}[f_1^\prime f_2^\prime f_1 f_2]^{1/2} \; 
\sum_{S^{\prime},S,i} 
\sum_{\substack{m_l, m_f \\ m_i,M_S}}\sum_{\substack{m_l^{\prime}, m_f^{\prime} \\ M_S^{\prime}}}
[S^{\prime}S]^{1/2} [i] \,
C^{f^\prime l^{\prime} J^{\prime}}_{m_f^{\prime},m_l^{\prime},m_J^{\prime}} C^{flJ}_{m_f,m_l,m_J} 
C^{S^{\prime}if^{\prime}}_{M_S^{\prime},m_i,m_f^{\prime}} C^{Sif}_{M_S,m_i,m_f} \nonumber  \\
&& \times  \Ninej{S_1^{\prime}}{S_2^{\prime}}{S^{\prime}}{i_1}{i_2}{i}
{f_1^\prime}{f_2^\prime}{f^{\prime}}
		\Ninej{S_1}{S_2}{S}{i_1}{i_2}{i}{f_1}{f_2}{f} 
		\langle l^{\prime},m_{l}^{\prime}|\langle (S_{1}^{\prime})_A,(S_{2}^{\prime})_B,S^{\prime},	M_{S}^{\prime}|	
									\hat{H}_\text{sd} |(S_{1})_A,(S_{2})_B;S,M_{S} \rangle	|l,m_{l}\rangle.			
\end{eqnarray}
After using (\ref{R23}) and (\ref{R24}) for the matrix elements of $\hat{H}_{\mathrm{sd}}$ in
the $\{|\alpha \rangle \}$ basis, the summations over magnetic quantum numbers can be reduced to 
summations over the three independent quantities $M_{S}$, 
$\epsilon = M_{S}^{\prime}-M_{S}$ and $\tau = m_{l}-M_{J}$ since
\begin{equation}
\label{A9}
M_{S}^{\prime}=M_{S}+\epsilon , \quad m_{f}=\tau, \quad m_{f}^{\prime}=\tau+\epsilon, 
\quad m_{l}=M_{J}-\tau, \quad
m_{l}^{\prime}=M_{J}-\tau -\epsilon , \quad m_{i}=\tau -M_{S}.
\end{equation}
These relationships give $M_{J}^{\prime}=m_{f}^{\prime}+m_{l}^{\prime}=m_{f}+m_{l}=M_{J}$. 
As only three Clebsch-Gordan coefficients now involve $M_{S}$, the summation over 
$M_{S}$ can be performed:
\begin{equation}
\label{A10}
\sum_{M_{S}} \Cleb{2}{S}{S^{\prime}}{\epsilon ,}{M_{S},}{M_{S}+\epsilon}
\Cleb{S^{\prime}}{i_{1}}{f^{\prime}}{M_{S}+\epsilon,}{\tau-M_{S},}{\tau+\epsilon}
\Cleb{S}{i_{1}}{f}{M_{S},}{\tau - M_{S},}{\tau} 
= (-1)^{-(f^{\prime}+i_{1}+S)}
[S^{\prime}f]^{1/2}
\Cleb{2}{f}{f^{\prime}}{\epsilon ,}{\tau ,}{\epsilon+\tau}
\Sixj{2}{S}{S^{\prime}}{i_{1}}{f^{\prime}}{f}.
\end{equation}
Similarly the summation over $\epsilon $ gives
\begin{equation}
\label{A11}
\sum_{\epsilon} 
\Cleb{f}{2}{f^{\prime}}{\tau ,}{\epsilon ,}{\epsilon+\tau}
\Cleb{f^{\prime}}{l^{\prime}}{J^{\prime}}{\epsilon+\tau ,}{M_{J}-\epsilon-\tau ,}{M_{J}}
\Cleb{2}{l^{\prime}}{l}{\epsilon ,}{M_{J}-\epsilon -\tau ,}{M_{J}-\tau}
=(-1)^{f+l^{\prime}+J^{\prime}}
[f^{\prime}l]^{1/2}
\Cleb{f}{l}{J^{\prime}}{\tau ,}{M_{J}-\tau ,}{M_{J}}
\Sixj{f}{2}{f^{\prime}}{l^{\prime}}{J^{\prime}}{l}.
\end{equation}
The remaining summation over $\tau $ gives $\delta_{J^{\prime},J}$
and the matrix element reduces to (\ref{R26b}).

For the 3--3 and 4--4 systems, the matrix elements
  $\langle a^\prime|\hat{H}_\mathrm{el} |a\rangle$ and
  $\langle a^\prime|\hat{H}_\mathrm{sd} |a\rangle$ are symmetrized combinations of
  $\langle a_{x^\prime y^\prime }^\prime | \hat{H}_\mathrm{el} | a_{xy}\rangle$ and
  $\langle a_{x^\prime y^\prime }^\prime | \hat{H}_\mathrm{sd} | a_{xy}\rangle$ respectively,
	where $x$ and $y$ can take the values 1 or 2. These
  combinations give rise to a selection rule $(-1)^{l-S-i}=1$, and two
  factors $\sqrt{2-\delta_{f_1 f_2}}$ and $\sqrt{2-\delta_{f_1^\prime f_2^\prime}}$. 
  The matrix elements both have the form
  \begin{equation}
\langle a^\prime|\hat{H}_z |a\rangle = \langle a_{12}^\prime | \hat{H}_z \hat{P}_{Si} 
\sqrt{(2-\delta_{f_1f_2})(2-\delta _{f_1^\prime f_2^\prime})} | a_{12}\rangle
\end{equation}
where $\hat{H}_z=\hat{H}_\mathrm{el}$ or $\hat{H}_\mathrm{sd}$, and $\hat{P}_{Si}$ is defined 
by its action on the $\{S,i\}$ basis:
\begin{equation}
\hat{P}_{Si} |S_1,S_2,S, M_S \rangle |i_1,i_2,i, m_i\rangle|l,m_l\rangle = \frac{1 + (-1)^{l-S-i}}{2}
|S_1,S_2,S, M_S \rangle |i_1,i_2,i, m_i\rangle|l,m_l\rangle
\end{equation}

\end{widetext}

\section{Integration of multichannel equations}

The multichannel equations (\ref{R10}) have been solved using two methods to
verify the numerical procedure using the Julia programming language
\cite{Julia}. The first method uses the renormalized Numerov
method~\cite{Johnson1978} on a linear grid of points consisting of connected
regions within which a fixed step size is used. The second method uses a
Runge-Kutta method with an adaptive step size to solve the equations (\ref{R10})
recast as first-order equations.

The solutions were found by integrating a linearly independent set of wave
functions outwards from $R=1\,a_{0}$ to $R=100\,a_{0}$ with the inner boundary
conditions $\mathbf{G}(R=1)=0$ and integrating a linearly independent set of
wave functions inwards from $R= 1000\,a_{0}$ to $R=100\;a_{0}$. The outer
boundary conditions were specified that all closed channels should be zero at
$R=1000\;a_0$. These two sets of solutions were matched to find a complete set
of allowed solutions (the number of solutions is the same as the number of open
channels) that satisfy both the inner and outer boundary conditions.

These solutions must then be matched to their asymptotic form (\ref{R34}) to
determine the S matrix. As the spin-dipole term decays slowly as $R^{-3}$, this
requires integration of the solutions (consisting only of open channels) to a
point well beyond $R=1000\,a_{0}$. We found that this integration is prone to
accumulated numerical error and so we chose to instead solve the integral
equations \cite{Joachain1983} for the coefficients of the asymptotic matching of
(\ref{R34}).  In this manner, the solutions $\mathbf{G}(R)$ were expressed in
the form
\begin{equation}
\label{B3}
\mathbf{G}(R) = \mathbf{A}(R) \mathbf{J}_{l_{a}}(k_{a}R) + 
\mathbf{B}(R) \mathbf{N}_{l_{a}}(k_{a}R).
\end{equation}
The matrices $\mathbf{A}(R)$ and $\mathbf{B}(R)$ satisfy the differential equations
\begin{eqnarray}
\label{B4}
\frac{d\mathbf{A}}{dR} &=& -\frac{1}{k_{a}}\mathbf{N}_{l_{a}}(k_{a}R)
\mathcal{U}(R)\mathbf{G}(R), \\
\frac{d\mathbf{B}}{dR} &=& \frac{1}{k_{a}}\mathbf{J}_{l_{a}}(k_{a}R)
\mathcal{U}(R)\mathbf{G}(R),
\end{eqnarray}
where
\begin{equation}
\label{B5}
\mathcal{U}_{a^{\prime}a}(R)=\frac{2 \mu}{\hbar^{2}}[V_{a^{\prime}a}(R) -
V_{aa}(R\rightarrow \infty)\delta_{a^{\prime},a}],
\end{equation}
and vary much more smoothly than the wave functions $\mathbf{G}(R)$.


\begin{thebibliography}{1}


\bibitem{Bloch2008} I.~Bloch, J.~Dalibard and W.~Zwerger,
  Many-body physics with ultracold gases,
  Rev. Mod. Phys. \textbf{80}, 885 (2008).
  
\bibitem{Tsubota2013} M.~Tsubota, M.~Kobayashi and H.~Takeuchi,
  Quantum hydrodynamics,
  Phys. Rep. \textbf{522}, 191 (2013).

\bibitem{Cooper2019} N.~R.~Cooper, J.~Dalibard and I.~B.~Spielman, 
  Topological bands for ultracold atoms,
  Rev. Mod. Phys. \textbf{91}, 015005 (2019).

\bibitem{Vassen2012} W.~Vassen, C.~Cohen-Tannoudji, M.~Leduc, D.~Boiron, C.~I.~Westbrook, 
A.~Truscott, K.~Baldwin, G.~Birkl, C.~Cancio, and M.~Trippenbach, 
Cold and trapped metastable noble gases,
Rev. Mod. Phys. \textbf{84}, 175 (2012)

\bibitem{Stas2004} R.~J.Stas, J.~M.~McNamara, W.~Hogervorst, and W.~Vassen,
Simultaneous magneto-optical trapping of a boson-fermion mixture of metastable helium atoms,
Phys. Rev. Lett. \textbf{93}, 053001 (2004)

\bibitem{Jeltes2007} T.~Jeltes, J.~M.~McNamara, W.~Hogervorst, W.~Vassen, V.~Krachmalnicoff, M.~Schellekens, A.~Perrin, H.~Chang, D.~Boiron, A.~Aspect and C.~I.~Westbrook
  Comparison of the Hanbury-Brown-Twiss effect for bosons and fermions,
  Nature \textbf{445}, 402 (2007).

\bibitem{Khakimov2016} R.~Khakimov, B.~Henson, D.~Shin, S.~Hodgman, R.~Dall,
  K.~Baldwin and A.~Truscott,
Ghost imaging with atoms,
Nature \textbf{540}, 100 (2016).

\bibitem{Henson2015} B.~M.~Henson, R.~I.~Khakimov, R.~G.~Dall, K.~G.~H.~Baldwin, L.-Y.~Tang and A.~G.~Truscott
  Precision Measurement for Metastable Helium Atoms of the 413 nm Tune-Out Wavelength at Which the Atomic Polarizability Vanishes
Phys. Rev. Lett. \textbf{115}, 043004 (2015).


\bibitem{Venturi2000} V.~Venturi and I.~B.~Whittingham, 
Close-coupled calculation of field-free collisions of cold metastable helium atoms,
Phys. Rev. A \textbf{61}, 060703(R) (2000)

\bibitem{Sirjean2002} O.~Sirjean, S.~Seidelin, J.~Vianna Gomes, D.~Boiron, C.~I.~Westbrook, 
A.~Aspect, and G.~V.~Shlyapnikov,
Ionization rates in a Bose-Einstein condensate of metastable helium,
Phys. Rev. Lett. \textbf{89}, 220406 (2002)

\bibitem{Mastwijk1998} H.~C.~Mastwijk, J.~W.~Thomsen, P.~van~der~Straten, and A.~Niehaus,
Optical collisions of cold metastable helium atoms,
Phys. Rev. Lett. \textbf{80}, 5516 (1998)

\bibitem{Tol1999} P.~J.~J.~Tol, N.~Herschbach, E.~A.~Hessels, W.~Hogervorst, and W.~Vassen,
Large numbers of cold metastable helium atoms in a magneto-optical trap,
Phys. Rev. A \textbf{60}, R761 (1999)

\bibitem{KM1999} M.~Kumakura and N.~Morita, Laser trapping of metastable ${}^{3}$He atoms: Isotopic
difference in cold Penning collisions,
Phys. Rev. Lett. \textbf{82}, 2848 (1999)


\bibitem{Stas2006} R.~J.~W.~Stas, J.~M.~McNamara, W.~Hogervorst, and W.~Vassen, Homonuclear
ionizing collisions of laser-cooled metastable helium atoms, 
Phys. Rev. A \textbf{73}, 032713 (2006)

\bibitem{Leo2001} P.~L.~Leo, V.~Venturi, I.~B.~Whittingham, and J.~F.Babb, Ultracold 
collisions of metastable helium atoms,
Phys. Rev. A \textbf{64}, 042710 (2001)


\bibitem{McNamara2007} J.~M.~McNamara, R.~J.~W.~Stas, W.~Hogervorst, and W.~Vassen,
Heteronuclear ionizing collisions between laser-cooled metastable helium atoms,
Phys. Rev. A \textbf{75}, 062715 (2007)

\bibitem{Dickinson2007} A.~S.~Dickinson, Quantum reflection model for ionization rate
coefficients in cold metastable helium collisions,
J. Phys. B: At. Mol. Opt. Phys \textbf{40}, F237 (2007)

\bibitem{Julienne1989} P.~S.~Julienne and F.~H.~Mies,
Collisions of ultracold trapped atoms,
J. Opt. Soc. Am. B \textbf{6}, 2257 (1989)

\bibitem{Cocks2015} D.~G.~Cocks, G.~Peach, and I.~B.~Whittingham,
Long-range states in excited ultracold  ${}^{3}$He$^{*}-{}^{4}$He$^{*}$ dimers,
J. Phys. B: At. Mol. Opt. Phys \textbf{48}, 115205 (2015)

\bibitem{Rosner1970} S.~D.~Rosner and F.~M.~Pipkin,
Hyperfine structure of the 2 ${}^{3}$S$_{1}$ state of He$^{3}$,
Phys. Rev. A \textbf{1}, 571 (1970)

\bibitem{Beams2006} T.~J.~Beams, G.~Peach, and I.~B.~Whittingham,
Spin-dipole-induced lifetime of the least-bound ${}^{5}\Sigma_{g}^{+}$ state of 
He$(2\,{}^{3}S_{1}) + $He$(2\,{}^{3}S_{1})$,
Phys. Rev. A \textbf{74}, 014702 (2006)

\bibitem{Julienne2002} P.~S.~Julienne, Ultracold collisions of Atoms and Molecules,
in \textit{Scattering: Scattering and Inverse Scattering in Pure and Applied Science},
editors P.~Sabatier and E.~R.~Pike (Academic Press, London, 2002), p. 1081

\bibitem{Garrison73} B.~J.~Garrison, W.~H.~Miller, and H.~F.~Schaffer,
Penning and associative ionization of triplet metastable helium atoms,
J. Chem. Phys. \textbf{59}, 3193 (1973)

\bibitem{Mies1980} F.~H.~Mies,
A scattering theory of diatomic molecules: General formalism using the channel state
representation,
Mol. Phys. \textbf{41}, 953 (1980)

\bibitem{Peach2017} G.~Peach, D.~G.~Cocks, and I.~B.~Whittingham,
Ultracold collisions in metastable helium,
J. Phys: Conf. Series \textbf{810}, 012003 (2017)

\bibitem{Seaton1966}  M.~J.~Seaton, 
Quantum defect theory I. General formulation,
Proc. Phys. Soc. \textbf{88}, 801 (1966)


\bibitem{Seaton1961}  M.~J.~Seaton, 
Strong coupling in optically allowed atomic transitions produced by electron impact,
Proc. Phys. Soc. \textbf{77}, 174 (1961)


\bibitem{Geltman1969} S.~Geltman, \textit{Topics in Atomic Collision Theory}
(Academic, New York, 1969), p. 183


\bibitem{Przy2005} M.~Przybytek and B.~Jeziorski,
Bounds for the scattering length of spin-polarized helium from high-accuracy electronic 
structure calculations,
J. Chem. Phys. \textbf{123}, 134315 (2005)

\bibitem{Muller1991} M.~W.~M\"{u}ller, A.~Merz, M.-W.~Ruf, H.~Hotop,  W.~Meyer, and
M.~Movre,
Experimental and theoretical studies of the Bi-excited collision systems 
He$^{*}(2\,{}^{3}S$)+ He$^{*}(2\,{}^{3}S, 2\,^{1}S)$ at thermal and subthermal 
kinetic energies,
Z. Phys. D: At. Mol. Clusters \textbf{21}, 89 (1991)

\bibitem{Mies2000} F.~H.~Mies and M.~Raoult,
Analysis of threshold effects in ultracold atomic collisions,
Phys. Rev. A \textbf{62}, 012708 (2000)

\bibitem{SM1994} J.~St\"{a}rck and W.~Meyer,
Long-range interaction potential of the ${}^{5}\Sigma^{+}_{g}$ state of He$_{2}$,
Chem. Phys. Lett. \textbf{225}, 229 (1994)

\bibitem{Browaeys2001} A.~Browaeys, A.~Robert, O.~Sirjean, J.~Poupard, S.~Nowak, D.~Boiron, 
C.~I.~Westbrook, and A.~Aspect,
Thermalization of magnetically trapped metastable helium,
Phys. Rev. A \textbf{64}, 034703 (2001)

\bibitem{Brink68} D.~M.~Brink and G.~R.~Satchler, \textit{Angular Momentum},
2nd ed. (Clarendon Press, Oxford, 1968)

\bibitem{Julia} J.~Bezanson, A.~Edelman, S.~Karpinski, and V.~B.~Shah,
Julia: A fresh approach to numerical computing,
Siam Rev. \textbf{59}, 65 (2017)

\bibitem{Johnson1978} B.~R.~Johnson,
The renormalized Numerov method applied to calculating bound states of the coupled-channel
Schroedinger equation,
J. Chem. Phys. \textbf{69}, 4678 (1978)

\bibitem{Cocks2010} D.~G.~Cocks, I.~B.~Whittingham and G.~Peach,
Effects of non-adiabatic and Coriolis couplings on the bound states of He(2${}^3$S)+He(2${}^3$P)
J. Phys. B: At. Mol. Opt. Phys. \textbf{43}, 135102 (2010).

\bibitem{Joachain1983} C.~J.~Joachain, \textit{Quantum Collision Theory}, third edition,
(North Holland, Amsterdam, 1983).

\end{thebibliography}
\end{document}